\documentclass[11pt]{article}

 \usepackage{graphics}
 \usepackage{graphicx}
 \usepackage{verbatim}
 \usepackage{color}
\usepackage{amssymb}
\usepackage{amsfonts}
\usepackage{amsmath}
\usepackage{graphicx}
\usepackage{caption}
\usepackage{subcaption}
\usepackage{latexsym}
\usepackage{dsfont}
\usepackage{color}
\usepackage{setspace}

\newcommand{\bb}{\begin{eqnarray}}
\newcommand{\ee}{\end{eqnarray}}
\newcommand{\beq}{\begin{equation}}
\newcommand{\eeq}{\end{equation}}
\newcommand{\ba}{\begin{array}}
\newcommand{\ea}{\end{array}}

\newtheorem{Remark}{Remark}

\newtheorem{Proposition}{Proposition}

\newtheorem{Definition}{Definition}

\title{Integrability and Linear Stability of Nonlinear Waves}

\author{Antonio Degasperis\\
Dipartimento di Fisica, ``Sapienza" Universit\`a di Roma, Rome, Italy \\
E-mail: antonio.degasperis@uniroma1.it
\and
Sara Lombardo\\
Department of Mathematics, Physics and Electrical Engineering\\
Northumbria University, Newcastle upon Tyne, UK\\
and\\
Department of Mathematical Sciences, School of Science\\
Loughborough University, Loughborough, UK\\
E-mail: s.lombardo@lboro.ac.uk
\and
Matteo Sommacal\\
Department of Mathematics, Physics and Electrical Engineering\\
Northumbria University, Newcastle upon Tyne, UK\\
E-mail: matteo.sommacal@northumbria.ac.uk}

\date{}

\DeclareCaptionLabelFormat{continued}{#1~#2 (Continued)}
\begin{document}

\bibliographystyle{plain}


\maketitle

\begin{abstract}
It is well known that the linear stability of solutions of $1+1$ partial differential equations which are integrable can be very efficiently investigated by means of spectral methods.

We present here a direct construction of the eigenmodes of the linearized equation which makes use only of the associated Lax pair with no reference to spectral data and boundary conditions. This local construction  is given in the general $N\times N$ matrix scheme so as to be applicable to a large class of integrable equations, including the multicomponent nonlinear Schr\"{o}dinger system and the multi-wave resonant interaction system.

The analytical and numerical computations involved in this general approach are detailed as an example for $N=3$ for the particular system of two coupled nonlinear Schr\"{o}dinger equations in the defocusing, focusing and mixed regimes. The instabilities of the continuous wave solutions are fully discussed in the entire parameter space of their amplitudes and wave numbers. By defining and computing the spectrum in the complex plane of the spectral variable, the  eigenfrequencies are explicitly expressed.  According to their topological properties, the complete classification of these spectra in the parameter space  is presented and graphically displayed. The continuous wave solutions are linearly unstable for a generic choice of the  coupling constants.

\vspace{1cm}

\vspace{0.3cm}
\end{abstract}

\newpage
\section{Introduction}
The problem of stability is central to the entire field of nonlinear wave propagation and is a fairly broad subject. Here, we are specifically concerned with the early stage of amplitude modulation instabilities due to quadratic and cubic nonlinearities, and we consider in particular dispersive propagation in a one-dimensional space, or diffraction in a two dimensional space.

After the first observations of wave instability \cite{BF1967,Rothenberg1990,Rothenberg1991} (see also \textit{e.g.} \cite{ZO2009}), the research on this subject has grown very rapidly because similar phenomena appear in various contexts such as water waves \cite{YL1980}, optics \cite{Agrawal1995}, Bose-Einstein condensation \cite{KFC2007} and plasma physics \cite{Kuznetsov1977}. Experimental findings were soon followed by theoretical and computational works. Predictions regarding short time evolution of small perturbations of the initial profile can be obtained by standard linear stability methods, see \textit{e.g.} \cite{SF1999} and references therein. Very schematically, if $u(x,t)$ is a particular solution of the wave equation, evolving in time $t$, and if $u+\delta u$ is the perturbed solution of the same equation, then, at the first order of approximation, $\delta u$ satisfies a linear equation whose coefficients depend on the solution $u(x,t)$ itself, and  therefore, they are generally non-constant. Consequently, solving the initial value problem $\delta u(x,0)\rightarrow \delta u(x,t)$ in general is not tractable by analytical methods. It is only for special solutions $u(x,t)$, such as nonlinear plane waves or solitary localized waves, see \textit{e.g.} \cite{SF1999}, that this initial value problem can be approached by solving an eigenvalue problem for an ordinary differential operator in the space variable $x$. In this way the computational task reduces to constructing the eigenmodes, \textit{i.e.} the eigenfunctions of an ordinary differential operator, while the corresponding eigenvalues are simply related to the proper frequencies. For very special solutions $u(x,t)$, this procedure exceptionally leads to a linearized equation with constant coefficients which can be solved therefore via Fourier analysis. A simple and well known example of this case is the linearization of the focusing nonlinear Schr{\"o}dinger (NLS) equation $iu_t + u_{xx} + 2|u|^2 u=0$ around its continuous wave (CW) solution  $u(x,t)= e^{2it}$. Here and thereafter, subscripts $x$ and $t$ denote partial differentiation, unless differently specified.
The computation of all the complex eigenfrequencies, in particular of their imaginary parts, yields the relevant information about the stability of $u(x,t)$, provided the set of eigenmodes be complete in the functional space characterized by the boundary conditions satisfied by the initial perturbation $\delta u(x,0)$. Since the main step of this method is that of finding the spectrum of a differential operator, the stability property of the solution $u(x,t)$ is also referred to as \emph{spectral stability}. It is clear that this method applies to a limited class of solutions of the wave equation. Although herein we are concerned with linear stability only, quite a number of studies on other forms of nonlinear waves stability have been produced by using different mathematical techniques and aimed at various physical applications. For instance, variational methods to assess orbital stability have been applied to solitary waves and standing waves, \textit{e.g.}, see \cite{MS1993} \cite{GO2012}.

An alternative and more powerful approach to stability originated from \cite{AKNS1974} shortly after the discovery of the complete integrability and of the spectral method to solve the Kortewg-de Vries (KdV) and NLS equations (\textit{e.g.}, see the textbooks \cite{AS1981,CD1982,NMPZ1984}). This method stems from the peculiar fact that the so-called \emph{squared eigenfunctions} (see next section for their definition) are solutions of the linearized equation solved by the perturbation $\delta u(x,t)$. Indeed, depending on boundary conditions, this technique yields a  representation of the perturbation $\delta u(x,t)$ in terms of such squared eigenfunctions. With respect to the spectral methods in use for non-integrable wave equations, the  squared eigenfunctions approach to stability shows its power by formally applying to almost any solution $u(x,t)$, namely also to cases where standard methods fail. Moreover this method, with appropriate algebraic conditions, proves to be applicable (see below Section \ref{sec:setup}) to a very large class of matrix Lax pairs and, therefore, to quite a number of integrable systems other than KdV and NLS equations (\textit{e.g.}, sine-Gordon, mKdV, derivative NLS, coupled NLS, three-wave resonant interaction, massive Thirring model, and other equations of interest in applications).
Its evident drawback is that its applicability is limited to the very special class of integrable wave equations. Notwithstanding this condition, it remains of important practical interest because several integrable partial differential equations have been derived in various physical contexts as reliable, though approximate, models \cite{dodd1982solitons, AS1981,dauxois2006physics}.
Moreover, the stability properties of particular solutions of an integrable wave equation provide a strong insight about similar solutions of a different non-integrable, but close enough, equation.
Among the many properties defining the concept of integrability the one that we consider here is the existence of a Lax pair of two linear ordinary differential equations for the same unknown, one in the space variable $x$ and the other in the time variable $t$ (see next Section), and whose compatibility condition is just the wave equation. Thus, a spectral problem with respect to the variable $x$ already appears at the very beginning of the integrability scheme.
With appropriate specifications, as stated below, this observation leads to the construction of the eigenmodes of the linearized equation, in terms of the solutions of the Lax pair. Moreover, via the construction of the  squared eigenfunctions one is able to compute the corresponding eigenfrequency $\omega$, which gives the (necessary and sufficient) information to assess linear  stability by the condition $\mathrm{Im}(\omega) >0$.
Explicit  expressions of such eigenmodes have been obtained if the unperturbed wave amplitude $u$ is a cnoidal wave (\textit{e.g.}  see \cite{KM1974,Sachs1983,KSF1984} for the KdV equation and \cite{KS1999} for the NLS equation), or if it is a soliton solution \cite{Yang2000} or, although only formally, an arbitrary solution \cite{Yang2002}. Therefore, the computational strategy amounts to constructing the set of eigenmodes and eigenfrequencies. It should be pointed out that the integrability methods, in an appropriate functional space of the wave fields $u(x,t)$, provide also the way of deriving the closure and completeness relations of the eigenmodes, see \textit{e.g.} \cite{Kaup1976,Yang2000,Yang2002} for solutions which vanish sufficiently fast as $|x|\rightarrow 0$. In this respect, a word of warning is appropriate. The boundary conditions imposed on the solutions $u(x,t)$ play a crucial role in proving that the wave evolution be indeed integrable. Thus, in particular for the NLS equation, integrability methods have been applied so far to linear stability of wave solutions which, as $|x|\rightarrow \infty$, either vanish as a localized soliton \cite{Yang2000}, or go to a CW solution (see the lecture notes \cite{DL2016}), or else are periodic, $u(x,t)=u(x+L,t)$ \cite{BDN2011}.
In these cases, by solving the so-called \emph{direct spectral problem}, to any solution $u(x,t)$ one can associate a set of spectral data, the spectral transform, say the analogue of the Fourier transform in a nonlinear context. This correspondence allows to formally solve the initial value problem of the wave equation. As a by-product, this formalism yields also a spectral representation of the small perturbations $\delta u(x,t)$ in terms of the corresponding small change of the spectral data. This connection is given by the squared eigenfunctions (see \cite{AKNS1974,Kaup1976,YK2009} for the NLS equation, and \cite{CD1982} for the KdV equation) which play the role which the Fourier exponentials have in the linear context. Indeed the squared eigenfunctions, which are computed by solving the Lax pair, are the eigenmodes of the linearized equation for $\delta u(x,t)$. This result follows from the inverse spectral transform machinery. However, as we show below, the squared eigenfunctions' property of being solutions of the linearized equation is a local one, as it  follows directly from the Lax pair without any need of the spectral transform.
More than this, integrability allows to go beyond the linear stage of the evolution of small perturbations. This is possible by the spectral method of solving the initial value problem for the perturbed solution $u+\delta u$ which therefore yields the long time evolution of $\delta u$ beyond the linear approximation, see for instance \cite{ZG2013,BM2016}. However, this important problem falls outside the scope of the present work and it will not be considered here (for the initial value problem and unstable solutions of the NLS equation, see \cite{GS2017a,GS2017b,GS2017c}).

The stability properties of a given solution $u(x,t)$ may depend on parameters. These parameters come from the coefficients of the wave equation, and from the parameters (if any) which characterize the solution $u(x,t)$  itself. This obvious observation implies that one may expect the parameter space to be divided into  regions where the solution $u(x,t)$ features different behaviours in terms of linear stability. Indeed this is the case, and crossing the border of one of these regions by varying the parameters, for instance a wave amplitude, may correspond to the opening of a gap in the instability frequency band,  so that a \emph{threshold} occurs at that amplitude value which corresponds to crossing.
The investigation of such thresholds is rather simple when dealing with scalar (one-component) waves. For instance, the KdV equation has no frozen coefficient, for a simple rescaling can set them equal to any real number, so that it reads $u_t+u_{xxx}+uu_x=0$. On the other hand, after rescaling, the NLS equation comes with a sign in front of the cubic term, distinguishing between defocusing and focusing self-interaction. These two different versions of the NLS equation lead to different phenomena such as modulation stability and instability of the continuous wave solution (for an introductory review, see \cite{DL2016}). Wave propagation equations which model different physical systems may have more structural coefficients whose values cannot be simultaneously, and independently, rescaled.
This is the case when two or more waves resonate and propagate while coupling to each other. In this case, the wave equations do not happen to be integrable for all choices of the coefficients. A well known example, which is the focus of Section \ref{sec:CNLS}, is that of two interacting fields, $u_j$, $j=1,\,2$, which evolve according to the coupled system of NLS equations
\begin{equation}
\label{expVNLS}
iu_{jt} +u_{jxx} -2\,(s_1|u_1|^2 + s_2|u_2|^2)\,u_j=0\,,\quad j=1,\,2\,,
\end{equation}
where $(s_1|u_1|^2 + s_2|u_2|^2)$ is the self- and cross-interaction term.
This is integrable only in three cases \cite{ZS1982}, namely (after appropriate rescaling): $s_1=s_2=1$ (defocusing Manakov model), $s_1=s_2=-1$ (focusing Manakov model) \cite{Manakov1973}, and the mixed case $s_1=-s_2=1$.
These three integrable systems of two coupled NLS (CNLS) equations are of interest in few special applications in optics \cite{Menyuk1987,EMGB1992,WM1999} and in fluid dynamics \cite{OPT2010}, while, in various contexts (\textit{e.g.} in optics \cite{Agrawal1995} and in fluid dynamics \cite{YL1980, AH2015}), the coupling constants $s_1$, $s_2$ take different values and the CNLS system happens to be non-integrable. Yet the analysis of the three integrable cases is still relevant in the study of the (sufficiently close) non-integrable ones \cite{YB1996}. The linear stability of CW solutions, $|u_j(x,t)|=$ constant, of integrable CNLS systems is of special interest not only because of its experimental observability, but also because it can be analysed via both standard methods and the squared eigenfunctions approach. As far as the standard methods are concerned, the linear stability of CW solutions has been investigated only for the focusing and defocusing regimes, but not for the mixed one ($s_1=-s_2$), and only in the integrable cases, by means of the Fourier transform \cite{FMMW2000}.
Conversely, as far as the integrability methods are concerned, it has been partially discussed in \cite{LZ2017} to mainly show that instability may occur also in defocusing media, in contrast to scalar waves which are modulationally unstable only in the focusing case.
In the following we approach the linear stability problem of the CW solutions of (\ref{expVNLS}) within the integrability framework to prove that the main object to be computed is a spectrum (to be defined below) as a curve in the complex plane of the spectral variable, together with the eigenmodes wave numbers and frequencies defined on it. In particular, we show that the spectrum which is relevant to our analysis is related to, but \emph{not coincident} with, the spectrum of the Lax equation for $\Psi$. In addition, if $\lambda$ is the spectral variable,  the computational outcome is the wave number $k(\lambda)$ and frequency $\omega(\lambda)$, so that the dispersion relation, and also the instability band, are implicitly defined over the spectrum through their dependence on $\lambda$. Since spectrum and eigenmodes depend on parameters, we explore the entire parameter space of the two amplitudes and coupling constants to arrive at a complete classification of spectra by means of numerically-assisted, algebraic-geometric techniques.
Our investigation in Section \ref{sec:CNLS} illustrates how the linear stability analysis works within the theory of integrability. Our focus is on $x$- and $t$-dependent CWs, a case which is both of relevance to physics and is computationally approachable. This case is intended to be an example of the general method developed in Section \ref{sec:setup}.
It is worth observing again that the linear stability of the CW solutions can indeed be discussed also by standard Fourier analysis, \textit{e.g.}, see \cite{FMMW2000} for the CNLS systems in the focussing and defocussing regimes. However, such analysis is of no help to investigate the stability of other solutions. On the contrary, at least for the integrable CNLS system (\ref{expVNLS}), our method relies only on the existence of a Lax pair and as such it has the advantage of being applicable also to other solutions as well. In particular, it can be applied to the CW solutions in all regimes (as we do it here), as well as to solutions such as, for instance, dark-dark, bright-dark and higher order solitons traveling on a CW background, to which the standard methods are not applicable.

This article is organized as follows. In the next section (Section \ref{sec:setup}) we give the general (squared eigenfunctions) approach together with the expression of the eigenmodes of the linearized equation for the  $N\times N$ matrix scheme, so as to capture a large class of integrable systems. There we define the $x$-spectrum in the complex plane of the spectral variable. In Section \ref{sec:CNLS} we provide an example of application of the theory by specializing the formalism introduced in Section \ref{sec:setup} to deal with the CNLS equations. We characterize the $x$-spectrum in the complex plane of the spectral variable according to their topological features, and we cover the entire parameter space according to five distinct classes of spectra. This characterization of the spectrum holds under the assumption that the small perturbation of the background CW solution is localized. Section \ref{sec:MI_CNLS} is devoted to discussing the classification of spectra and the corresponding stability features in the focusing, defocusing and mixed coupling regimes, in terms of the physical parameters while a conclusion with open problems and perspectives of future work is the content of Section \ref{sec:conclusion}. Details regarding computational and numerical aspects of the problem are confined in Appendices.

\section{Integrable wave equations and small perturbations}
\label{sec:setup}
The integrable partial differential equations (PDEs) which are considered here are associated with the following pair of matrix ordinary differential equations (ODEs), also known as \emph{Lax pair} (\textit{e.g.}, see \cite{CD1982,NMPZ1984,AC1991}),
\begin{equation}
\label{laxpair}
\Psi_x=X\Psi\,, \quad\Psi_t=T\Psi\,,
\end{equation}
where $\Psi$, $X$ and $T$ are $N\times N$ matrix-valued complex functions. The existence of a fundamental (\textit{i.e.} non singular) matrix solution $\Psi=\Psi(x,t)$ of this overdetermined system is guaranteed by the condition that the two matrices $X$ and $T$ satisfy the differential equation
\begin{equation}
\label{compat}
X_t - T_x + [X\,,\,T] =0\,.
\end{equation}
We recall here that, unless differently specified,
a subscripted variable means partial differentiation with respect to that variable, and $[ A \,,\, B ]$ stands for the commutator $ AB-BA$.
In order to identify this condition (\ref{compat}) as an integrable partial differential equation for some of the entries of the matrix $X$, it is essential that both matrices $X$ and $T$ parametrically depend on an additional complex variable $\lambda$, known as the \emph{spectral parameter}.
In order to make this introductory presentation as simple as possible, we assume that $X(\lambda)$ and $T(\lambda)$  be polynomial in $\lambda$ with degrees $n$ and $m$, respectively.
As a consequence, the matrix $X_t - T_x + [X\,,\,T]$ is as well polynomial in $\lambda$ with degree $n+m$ and therefore the compatibility equation (\ref{compat}) yields  $n+m+1$  equations for the matrix coefficients of the polynomials $X$ and $T$.

If the pair $X$ and $T$ is a given solution of (\ref{compat}), we consider a new solution $X\rightarrow X+\delta X$, $T\rightarrow T+\delta T$, which differs by a small change of  the matrices $X$ and $T$, with the implication that the pair of matrices $\delta X$ and $\delta T$, at the first order in this small change, satisfies the \emph{linearized equation}
\begin{equation}
\label{linearcompat}
(\delta X)_t - (\delta T)_x + [\delta X\,,\,T] + [X\,,\,\delta T] =0\,.
\end{equation}
Again the left-hand side of this linearized equation has a polynomial dependence on $\lambda$ and the vanishing of all its coefficients results in a number of algebraic or differential equations. These obvious observations lead us to focus on the matrix linearized equation (\ref{linearcompat}) itself, which reads,
by setting $A= \delta X\,,\,B=\delta T$,
\begin{equation}\label{LE}
A_t - B_x + [A\,,\,T] + [X\,,\,B] =0\,,
\end{equation}
and to search for its solutions $A(x,t,\lambda)$ and $B(x,t,\lambda)$.
Our main target is to find those solutions  which are related to the fundamental matrix solution $\Psi(x,t,\lambda)$ of the Lax pair (\ref{laxpair}).
To this purpose we first note that the similarity transformation
\begin{equation}\label{phi}
M(\lambda) \rightarrow \Phi(x,t,\lambda)=\Psi(x,t,\lambda)\, M(\lambda)\, \Psi^{-1}(x,t,\lambda)\,,
\end{equation}
of a constant (\textit{i.e.} $x$, $t$-independent) matrix $M(\lambda)$ yields the transformed matrix $\Phi$ which, for any given arbitrary matrix $M$, satisfies the pair of linear ODEs
\begin{equation}
\label{phipair}
\Phi_x=[X,\Phi]\,,\quad \Phi_t=[T,\Phi]\,.
\end{equation}
Equations (\ref{phipair}) are compatible with each other because of (\ref{compat}). Then, for future reference, we point out the following observations.
\begin{Proposition}
\label{prop:1}
If the pair $A\,,\,B$ solves the linearized equation (\ref{LE}) then also the pair
\begin{equation}\label{newsolut}
F=[A\,,\,\Phi]\,,\quad G=[B\,,\,\Phi]
\end{equation}
 is a solution of the same linearized equation (\ref{LE}), namely
\begin{equation}\label{FGLE}
F_t - G_x + [F\,,\,T] + [X\,,\,G] =0\;\;.
\end{equation}
\end{Proposition}
This is a straight consequence of the Jacobi identity and of the assumption that the matrix $\Phi$ be a solution of (\ref{phipair}).

\begin{Proposition}
\label{prop:2}
The following expressions
 \begin{equation}\label{solutionLE}
F=c\left[\frac{\partial X}{\partial \lambda}\;,\;\Phi\right]\,,\quad G=c\left[\frac{\partial T}{\partial \lambda}\;,\;\Phi\right]
\end{equation}
provide a  solution of the linearized equation (\ref{FGLE}), $c$ being an arbitrary complex number. 
\end{Proposition}
The validity of this statement follows from the fact that the matrices
\begin{equation}
\label{ABsolution}
A=\frac{\partial X }{ \partial \lambda}\,,\quad B=\frac{\partial T }{ \partial \lambda}\,,
 \end{equation}
obviously solve the linearized equation (\ref{LE}), and from Proposition \ref{prop:1}.

 At this point we go back to the nonlinear matrix PDE which follows from the condition (\ref{compat}).
In view of the applications within the theory of nonlinear resonant phenomena that we have in mind (which include the multicomponent nonlinear Schr\"{o}dinger system and the multi-wave resonant interaction system), we assume that the polynomials $ X(\lambda)$ and $T(\lambda)$, see (\ref{laxpair}),  have degree one and, respectively, two, namely
\begin{equation}\label{XTpair}
X(\lambda)=i \lambda \Sigma + Q\,,\quad T(\lambda)=\lambda^2 T_2 + \lambda T_1 + T_0\,.
\end{equation}
where $\Sigma$, $Q$, $T_0$, $T_1$ and $T_2$ are matrix-valued functions of $x$ and $t$.
The extension to higher degree polynomials results only in an increased computational effort.

Moreover, before  proceeding further, few preliminary observations and  technicalities are required. First, we assume that the $N\times N$ matrix $\Sigma$, see (\ref{XTpair}), be constant and Hermitian. Therefore, without any loss of generality, $\Sigma$ is set to be diagonal and real, namely, in block-diagonal notation,
\begin{equation}\label{matrixSigma}
\Sigma=\textrm{diag}\{\alpha_1 \mathds{1}_1,\dots, \alpha_L \mathds{1}_L \}\,,\quad 2 \leq L \leq N\,,
\end{equation}
where the real eigenvalues $\alpha_j$, $j=1,\dots , L$, satisfy $\alpha_j \neq \alpha_k$ if $j\neq k$, while $\mathds{1}_j $ is the $n_j \times n_j$ identity matrix where $n_j$ is the (algebraic) multiplicity of the eigenvalue $\alpha_j$.
Of course, $\sum_{j=1}^L n_j =N$. Note that this matrix $\Sigma$ induces the splitting of the set of $N\times N$ matrices into two subspaces, namely that of \emph{block-diagonal} matrices and that of \emph{block-off-diagonal} matrices. Precisely, if $\mathcal{M}$ is any $N \times N$ matrix, then we adopt the following notation
\begin{equation}\label{matrixsplit}
\mathcal{M}=\mathcal{M}^{(d)} + \mathcal{M}^{(o)},
\end{equation}
where
\begin{subequations}
\label{matrixblock}
\begin{equation}
\label{blockdiag}
\mathcal{M}^{(d)} = \left ( \begin{array} {ccccc}  \framebox[1.5cm][c]{$n_1 \times n_1$} & 0 & 0 & 0 & 0 \\ 0 & \framebox[1.5cm][c]{$\cdot$} & 0 & 0 & 0 \\ 0 & 0 & \framebox[1.5cm][c]{$\cdot$} & 0 & 0 \\ 0 & 0 & 0 & \framebox[1.5cm][c]{$\cdot$} & 0 \\ 0 & 0 & 0 & 0 & \framebox[1.5cm][c]{$n_L \times n_L$} \end{array} \right )
\end{equation}
is the block-diagonal part of $\mathcal{M}$, and
\begin{equation}\label{blockoffdiag}
\mathcal{M}^{(o)} = \left ( \begin{array} {ccccc} 0 & \framebox[1.5cm][c]{$n_1 \times n_2$} & \framebox[1.5cm][c]{$\cdot$} & \framebox[1.5cm][c]{$\cdot$} & \framebox[1.5cm][c]{$n_1\times n_L$}  \\ \framebox[1.5cm][c]{$n_2 \times n_1$} & 0 & \framebox[1.5cm][c]{$\cdot$} & \framebox[1.5cm][c]{$\cdot$} & \framebox[1.5cm][c]{$\cdot$}  \\ \framebox[1.5cm][c]{$\cdot$} & \framebox[1.5cm][c]{$\cdot$} & 0 & \framebox[1.5cm][c]{$\cdot$} & \framebox[1.5cm][c]{$\cdot$} \\ \framebox[1.5cm][c]{$\cdot$} & \framebox[1.5cm][c]{$\cdot$} & \framebox[1.5cm][c]{$\cdot$} & 0 & \framebox[1.5cm][c]{$\cdot$}   \\ \framebox[1.5cm][c]{$n_L \times n_1$} & \framebox[1.5cm][c]{$\cdot$} & \framebox[1.5cm][c]{$\cdot$} & \framebox[1.5cm][c]{$\cdot$} & 0 \end{array} \right )
\end{equation}
\end{subequations}
is its block-off-diagonal part. Consistently with this notation, the entries $\mathcal{M}_{jk}$ of an $N\times N$ matrix $\mathcal{M}$ are meant to be matrices themselves of dimension $n_j \times n_k$ with the implication
that the ``matrix elements'' $\mathcal{M}_{jk}$ may not commute with each other (\textit{i.e.} the subalgebra of block-diagonal matrices, see (\ref{blockdiag}), is non-commutative). Moreover, one should keep in mind that the off-diagonal entries $\mathcal{M}_{jk}$ are generically rectangular. We also note that the product of two generic $N\times N$ matrices $\mathcal{A}$ and $\mathcal{B}$ follows the rules: $\left (\mathcal{A}^{(d)} \,\mathcal{B}^{(d)} \right )^{(o)}=0$, $\left(\mathcal{A}^{(d)} \,\mathcal{B}^{(o)} \right )^{(d)}=0$ while, for $N>2$, the product $\mathcal{A}^{(o)} \,\mathcal{B}^{(o)}$ is neither block-diagonal nor block-off-diagonal.

Next the matrix $Q(x,t)$ in (\ref{XTpair}) is taken to be  block-off-diagonal, whose entries are assumed to be functions of $x$, $t$ only. Its required property  is just differentiability up to sufficiently high order while no relation among its entries is assumed.

The matrix $T$, see (\ref{XTpair}), satisfies the compatibility condition (\ref{compat}), which entails the following expression of the coefficients
\begin{equation}\label{Tmatrix}
\begin{array}{lll}
T_2 =& \!\!C_2 \,,\\
T_1 = & \!\!C_1 -i\, I_1 -i\, D_2( Q )\,, \\
T_0 = & \!\! C_0 + I_0  -\frac12 [D_2(Q)\,,\,\Gamma(Q)]^{(d)} +\\
& \!\!- \Gamma(D_2( Q_x )) - \Gamma([ D_2(Q)\,,\,Q ]^{(o)})-i\, D_1(Q) - [ I_1\,,\, \Gamma(Q) ] \,,
\end{array}
\end{equation}
with  the following comments and definitions. The matrices $C_j$, $j=0, 1, 2$, are constant and block-diagonal, $C_j^{(o)}=0$. In the following we set $C_0=0$, because this matrix is irrelevant to our purposes. The notation $\Gamma (\cdot)$ stands for the linear invertible map acting only on the subspace of the block-off-diagonal matrices (\ref{blockoffdiag}) according to the following definition and properties
\begin{equation}
\label{gamma}
\left(\Gamma(\mathcal{M})\right)_{jk} = \frac{\mathcal{M}_{jk}}{\alpha_j-\alpha_k}\,,\qquad[\Sigma \,,\, \Gamma(\mathcal{M})] =   \Gamma([\Sigma \,,\, \mathcal{M}]) = \mathcal{M},\qquad \mathcal{M}^{(d)} =0\,,
\end{equation}
which show that also the matrix $\Gamma(\mathcal{M})$ is  block-off-diagonal. Also the maps $D_j (\cdot)$, $j=1, 2$, act only on block-off-diagonal matrices according to the definitions
\begin{equation}\label{D1D2}
D_j(\mathcal{M})  = [ C_j\,,\,\Gamma(\mathcal{M})] =\Gamma([ C_j\,,\,\mathcal{M}]) \,,
\,,\qquad \mathcal{M}^{(d)} =0\,,\quad j =1,2\,.
\end{equation}
Finally, the matrices $I_j$, $j=1, 0$, are block-diagonal and take the integral expression
\begin{subequations}
\label{integ}
\begin{align}
I_1(x,t) =& \int^x \mathrm{d}y [ Q(y,t)\,,\,D_2(Q(y,t))]^{(d)}\,,\\
\nonumber\\
I_0(x,t) =& \int^x \mathrm{d}y \, \bigg\{-\frac{1}{2} \left[C_2\,,\,[\Gamma(Q_y(y,t))\,,\,\Gamma(Q(y,t)) ]^{(d)}\right] +\nonumber\\
& \hspace{1.3cm}-\left[ Q(y,t)\,,\, \Gamma( [D_2(Q(y,t))\,,\,Q(y,t)]^{(o)}) \right]^{(d)} +\nonumber\\
&  \hspace{1.3cm}-i\,\Big[Q(y,t)\,,\,D_1(Q(y,t))\Big]^{(d)} - \Big[Q(y,t)\,,\,[ I_1(t)\,,\,\Gamma(Q(y,t))]\,\Big]^{(d)} \bigg\}\,.
\end{align}
\end{subequations}
Because of (\ref{integ}), the matrix $Q(x,t)$ evidently satisfies an  integro-differential, rather than a partial differential, equation as a consequence of  the compatibility condition (\ref{compat}). Incidentally, we note that this kind of non-locality  generated by the Lax pair (\ref{laxpair}) has been already pointed out \cite{degasperis2011}, while its solvability via  spectral methods remains an open question. Here, however, we focus our attention on local equations only. Therefore the condition that $Q(x,t)$ evolves  according to a partial differential equation is equivalent to the vanishing of the matrices $I_j$, $j=1, 0$, which ultimately implies restrictions on the constant matrices $C_1$ and $C_2$, as specified by the following
\begin{Proposition}
\label{pro:3}
The matrices $I_1$ and $I_0$ identically  vanish if, and only if,
 the blocks of $C_1$ and $C_2$ are proportional to the identity matrix, namely
\begin{equation}
\label{matrixC}
C_1=\mathrm{diag}\{\beta_1 \mathds{1}_1,\dots, \beta_L \mathds{1}_L \}\,, \quad
C_2=\mathrm{diag}\{\gamma_1 \mathds{1}_1,\dots, \gamma_L \mathds{1}_L \} \,.
\end{equation}
\end{Proposition}
Through the rest of the paper, we maintain these locality conditions so that the resulting evolution equation for the matrix $Q$ reads
\begin{align}
\label{integPDE}
Q_t =&\; -\Gamma(D_2(Q_{xx})) -[\Gamma(D_2(Q_x))\,,\,Q]^{(o)} - \Gamma([D_2(Q))\,,\,Q]_x^{(o)}) + \nonumber\\
& - [ (D_2(Q) \Gamma(Q))^{(d)}\,,\,Q]  \nonumber\\
& \!-[ \Gamma([D_2(Q)\,,\,Q]^{(o)})\,,\,Q]^{(o)} -i\, D_1(Q_x)  -i\, [  D_1(Q)\,,\,Q] ^{(o)} \,.
\end{align}
This equation can be linearized around a given solution $Q(x,t)$ by substituting  $Q$ with $Q+\delta Q$ and by neglecting all nonlinear terms in the variable $\delta Q$. In this way we obtain the following linear PDE
\begin{align}
\label{lineardelta}
\delta Q_t  =& \; -\Gamma(D_2(\delta Q_{xx})) - [\Gamma(D_2(\delta Q_x))\,,\,Q]^{(o)} -[\Gamma([D_2( Q_x)\,,\,\delta Q]^{(o)} + \nonumber\\
& -\Gamma([D_2(\,\delta Q)\,,\, Q]^{(o)}+[D_2(Q)\,,\,\delta Q]^{(o)})_x  -[\left (D_2(Q) \,\Gamma(Q)\right)^{(d)}\,,\,\delta Q] + \nonumber\\
& - [\, ( D_2( \delta Q)\,\Gamma(Q) )^{(d)} \,,\, Q ] - [\, ( D_2( Q)\,\Gamma(\delta Q) )^{(d)} \,,\, Q ] +\nonumber\\
& - [ \Gamma([D_2(Q)\,,\,Q]^{(o)})\,,\,\delta Q]^{(o)}- [ \Gamma([D_2(\delta Q)\,,\,Q]^{(o)})\,,\,Q]^{(o)}  +\nonumber\\
&- [ \Gamma([D_2(Q)\,,\,\delta Q]^{(o)})\,,\,Q]^{(o)} -i\,D_1 (\delta Q_x) +\nonumber\\
& -i\, [D_1 (Q)\,,\,\delta Q ]^{(o)} -i\,  [D_1 (\delta Q)\,,\,Q ]^{(o)} \, .
\end{align}
 We are now in the position to formulate our next proposition which is the main result of this section. \\
\begin{Proposition}
\label{prop:4}
The matrix
\begin{equation}\label{Fmatrix}
F= [\Sigma\,,\, \Phi ]\,,
\end{equation}
defined by (\ref{solutionLE}), together with (\ref{XTpair}), satisfies the same linear PDE (\ref{lineardelta}) satisfied by $\delta Q$ if and only if the block-diagonal matrices $C_1$ and $C_2$ satisfy the  same conditions (\ref{matrixC}) which guarantee the local character of the evolution equation (\ref{integPDE}).
 \end{Proposition}

\textbf{Sketch of a proof}\hspace{0.3cm} The proof of this result is obtained by straight computation, which is rather long and tedious. Therefore we skip detailing plain steps, but, for the benefit of the reader, we point out here just a few hints on how to go through the calculation which is merely algebraic. \\
The guiding idea is to eliminate the variable $\lambda$ while combining the two ODEs (\ref{phipair}) so as to obtain a PDE with $\lambda$-independent coefficients for the matrix $F= [\Sigma\,,\,\Phi]$. To this purpose, one starts by rewriting the first of the ODEs (\ref{phipair}) for the block-diagonal $\Phi^{(d)}$ and block-off-diagonal $\Phi^{(o)} = \Gamma (F)$ components of $\Phi$, namely
\begin{equation}\label{FRELAT}
\lambda F= \Gamma(F_x) - [Q\,,\,\Phi^{(d)}] - [Q\,,\,\Gamma(F)]^{(o)}\,,\qquad \Phi^{(d)}_x = [Q\,,\,\Gamma(F) ]^{(d)} \,.
\end{equation}
Next, one considers the time derivative $F_t$ by using expression (\ref{Fmatrix}), along with the right-hand-side commutator of the second relation in (\ref{phipair}), namely $F_t=[\Sigma\,,\,[T\,,\, \Phi] ]$, and one replaces the matrix $T(\lambda)$ with its expression as in (\ref{XTpair}) and (\ref{Tmatrix}). Then, all terms of the form $\lambda F$, wherever they appear, should be replaced by the right-hand-side expression given in the the first equation in (\ref{FRELAT}), in order to eliminate the spectral variable $\lambda$. The outcome of this substitution is that all terms containing the matrix $\Phi^{(d)}$ do indeed cancel out, provided conditions (\ref{matrixC}) are satisfied. The remaining terms of the expression of $F_t$ can be rearranged as to coincide with those which appear in the right-hand-side expression of the linearized PDE (\ref{lineardelta}), by making use of algebraic matrix identities. \hfill$\square$

Few remarks are in order to illustrate these findings. The matrix
\begin{equation}\label{FPsi}
F= [\Sigma\,,\,\Psi M \Psi^{-1}]
\end{equation}
(see (\ref{Fmatrix}) and (\ref{phi})), which has the same block-off-diagonal structure (\ref{blockoffdiag}) of the matrix $\delta Q$, turns out to be a $\lambda$-dependent solution of the linearized equation (\ref{lineardelta}). Its $\lambda$-dependence originates possibly from the (still arbitrary) matrix $M$, and certainly from the matrix solution $\Psi$ of the Lax pair (\ref{laxpair}). Indeed, $F$ plays the same role as the exponential solution of linear equations with constant coefficients, namely, by varying $\lambda$ over a spectrum (see below), it provides the set of the \emph{Fourier-like} modes of the linear PDE (\ref{lineardelta}). 
In this respect we note that the property of the matrix $F$ of satisfying the linearized PDE (\ref{lineardelta}) does not depend on the boundary values of $Q(x,t)$ at $x=\pm \infty$. In other words, this result applies as well to solutions $Q(x,t)$ of the integrable PDE (\ref{integPDE}) with vanishing and non-vanishing boundary values, or to periodic solutions, as required in various physical applications. This statement follows from the fact that Proposition \ref{prop:4} has been obtained by algebra and differentiation only. It is however clear that the boundary conditions affect the expression of the matrix $F$ through its definition (\ref{FPsi}) in terms of the solution $\Psi$ of the Lax pair of ODEs (\ref{laxpair}).
Moreover, with appropriate reductions imposed on the matrix $Q(x,t)$,  the integrable matrix PDE (\ref{integPDE}) associated with the Lax pair (\ref{laxpair}) with (\ref{XTpair}) includes, as special cases, wave propagation equations of physical relevance. Indeed, the PDE corresponding to $L=2$, $N=2$ and $C_1=0$ yields the nonlinear Schr\"{o}dinger equation, while for its multicomponent versions (such as the vector or matrix generalizations of the NLS equation) one may set $L=2$, $N>2$ and $C_1=0$. By setting instead $L=N>2$ and $C_2=0$, one obtains the three wave resonant interaction system for $N=3$ \cite{K1976}, and many-wave interaction type equations for $N>3$  \cite{AS1981} (\textit{e.g.}, see \cite{CD2004,DL2007,DL2009}).

\subsection{$x$-Spectrum $\textbf{S}_x$ of the solution $Q(x, t)$}
\label{subsec:spectrum}
The result stated by Proposition \ref{prop:4} in the previous section implies that any sum and/or integral of $F(x,t,\lambda)$ over the spectral variable $\lambda$  is a solution $\delta Q$ of the linear equation (\ref{lineardelta}). Here and in the following, we assume that such linear combination of matrices $F(x,t,\lambda)$, which formally looks as the Fourier-like integral representation
\begin{equation}\label{lincomb}
\delta Q(x,t)=\int \mathrm{d}\lambda \,F(x,t,\lambda)
\end{equation}
is such to yield a solution $\delta Q$ of (\ref{lineardelta}) which is both \emph{bounded} and \emph{localized} in the $x$ variable at any fixed time $t$ (periodic perturbations $\delta Q(x, t)$ are not considered here). The boundedness of $\delta Q$ implies that the matrix $F(x,t,\lambda)$ be itself bounded on the entire $x$-axis for any fixed value of $t$ and for any of the values of $\lambda$ which appear in integral (\ref{lincomb}).  This boundedness condition of $F(x,t,\lambda)$ defines a  special subset of the complex $\lambda$-plane, over which the integration runs, which will be referred to as the \emph{$x$-spectrum} $\textbf{S}_x$ of the solution $Q(x, t)$. This spectrum obviously depends on the behaviour of the matrix $Q(x)$ for large $|x|$. Indeed, if $Q(x, t)$ vanishes sufficiently fast as $|x|\rightarrow\infty$ (like, for instance, when its entries are in $L^1$), then $\textbf{S}_x$ coincides with the spectrum of the differential operator $\mathrm{d}/\mathrm{d}x -i\lambda\,\Sigma - Q(x)$ defined by the ODE $\Psi_x=X\Psi$ (\textit{i.e.}, the $x$-part of the Lax pair (\ref{laxpair})). However, if instead $Q(x, t)$ goes to a non-vanishing and finite value as $|x|\rightarrow\infty$, this being the case for continuous waves, the spectrum $\textbf{S}_x$ associated with the solution $Q(x, t)$ \emph{does not} coincide generically with the spectrum of the differential operator $\mathrm{d}/\mathrm{d}x -i\,\lambda\,\Sigma - Q(x)$. In the present matrix formalism, this happens for $N>2$, and it is due to the fact that the spectral analysis applies to the ODE $\Phi_x=[X,\Phi]$, see (\ref{phipair}), rather than to the Lax equation $\Psi_x=X\Psi$ itself. We illustrate this feature for $N=3$ in the next section, where we consider the stability of CW solutions.
The spectrum $\textbf{S}_x$ consists of a piecewise continuous curve and possibly of a finite number of isolated points. Note that at any point $\lambda$ of the spectrum $\textbf{S}_x$ the matrices $F(x,t,\lambda)$, for any fixed $t$, span a functional space of matrices whose dimension depends on $\lambda$.
Here, we do not take up the issue of the completeness and closure of the set of matrices $F(x,t,\lambda)$ and we rather consider solutions of (\ref{lineardelta}) which have the integral representation (\ref{lincomb}) and vanish sufficiently fast as $|x|\rightarrow\infty$, this being a requirement on the matrix $M(\lambda)$ which appears in definition (\ref{Fmatrix}) with (\ref{phi}).
As in the standard linear stability analysis, the given solution  $Q(x, t)$ is then linearly stable if any initially small change  $\delta Q(x, t_0)$ remains small as time grows, say $t>t_0$. Thus, the basic ingredient of our stability analysis is the time-dependence of the matrices $F(x,t,\lambda)$ for any $\lambda \in \textbf{S}_x$.
As we show below, getting this information requires knowing the spectrum $\textbf{S}_x$ whose computation therefore is our main goal.

\section{Wave coupling and spectra: an example}
\label{sec:CNLS}

In order to illustrate the results of the previous section and to capture, at the same time, a  class of nonlinear wave phenomena of physical relevance (including the system of two coupled NLS equations), the rest of the paper will focus on the matrix case $N=3$, $L=2$, with
\begin{equation}\label{matrixsigma}
\Sigma=\textrm{diag}\{1 \,,\, -1\,,\, -1 \}
\end{equation}
and
\begin{equation}\label{matrixQ}
Q=\left (\begin{array}{ccc} 0 & v_1^* & v_2^* \\ u_1 & 0 & 0 \\ u_2 & 0 & 0 \end{array} \right )\,.
\end{equation}
Here and below the asterisk  denotes complex conjugation and the four field variables $u_1,u_2,v_1,v_2$ are considered as independent functions of $x$ and $t$, and are conveniently arranged as two two-dimensional vectors, that is
\begin{equation}\label{vectoru-v}
 \mathbf{u}=\left (\begin{array}{c} u_1 \\ u_2  \end{array} \right )\,,\quad   \mathbf{v}=\left (\begin{array}{c} v_1 \\ v_2  \end{array} \right )\,.
\end{equation}
While we conveniently stick to this notation in this section, these formulae will be eventually reduced to those which lead to two coupled NLS equations as discussed in Section \ref{sec:MI_CNLS}. Indeed, we note that adopting the non-reduced formalism (\ref{matrixQ}), as we do it here, makes this presentation somehow simpler. In the present case, all matrices are $3\times 3$, and the $T$-matrix (\ref{Tmatrix}), with $C_2=2i\Sigma$, $C_1=C_0=0$, specializes to the expression
\begin{equation}\label{3x3Tmatrix}
 T(\lambda)=2i \lambda^2 \Sigma + 2\lambda Q + i \Sigma(Q^2-Q_x)\,.
\end{equation}
Then the matrix PDE (\ref{integPDE}) becomes
\begin{equation}\label{Qeq}
 Q_t=-i \,\Sigma (Q_{xx} -2Q^3)\,,
\end{equation}
which is equivalent to the two vector PDEs
\begin{align}
\label{2vecteq}
\mathbf{u}_t &=i[ \mathbf{u}_{xx} -2( \mathbf{v}^{\dagger}  \mathbf{u})  \mathbf{u}]\,\nonumber\\
\mathbf{v}_t &=i[ \mathbf{v}_{xx} -2( \mathbf{u}^{\dagger}  \mathbf{v})  \mathbf{v}]\,.
\end{align}
Here, the dagger notation denotes the Hermitian conjugation (which takes column-vectors into row-vectors).
In this simpler setting, if $Q(x,t)$ is a given solution of the equation (\ref {Qeq}), the linearized equation (\ref{lineardelta}) for a small change $\delta Q(x,t)$ reads
\begin{equation}\label{linQeq}
 \delta Q_t=-i \,\Sigma [\delta Q_{xx} -2(\delta Q Q^2 +Q \delta Q Q + Q^2 \delta Q)] \,.
\end{equation}
Moreover, Proposition \ref{prop:4} guarantees that the matrix $F(x, t, \lambda)$ satisfies  this same linear PDE, namely
\begin{equation}\label{Feq}
 F_t=-i \,\Sigma [F_{xx} -2(F Q^2 +Q F Q + Q^2 F)]\,\,,
\end{equation}
and, for $\lambda \in \textbf{S}_x$, these solutions should be considered as eigenmodes of the linearized equation.

The spectral analysis based on Proposition \ref{prop:4} applies to a large class of solutions $Q(x,t)$ of the nonlinear wave equation (\ref{Qeq}).  However, analytic computations are achievable if the fundamental matrix solution $\Psi(x, t, \lambda)$ of the Lax pair corresponding to the solution $Q(x,t)$ is explicitly known. Examples of explicit solutions of the Lax pair are known for particular $Q(x,t)$, for instance multisoliton (reflectionless) solutions (see the stability analysis in \cite{K2007}), cnoidal waves, and continuous waves.
Here we devote our attention to the stability of the periodic, continuous wave (CW) solution of (\ref{Qeq}), or of the equivalent vector system (\ref{2vecteq}),
\begin{equation}
\label{2vectsolut}
\mathbf{u}(x,t)=e^{i(qx\sigma_3-\nu t)}  \mathbf{a}\,,\quad  \mathbf{v}(x,t)=e^{i(qx\sigma_3-\nu t)}  \mathbf{b}\,,\quad \nu=q^2+2 \mathbf{b}^{\dagger}  \mathbf{a} \,.
\end{equation}
In these expressions $a$ and $b$ are arbitrary, constant and, with no loss of generality, real 2-dim vectors:
\begin{equation}
\label{vectora-b}
 \mathbf{a}=\left (\begin{array}{c} a_1 \\ a_2  \end{array} \right ) \,,\quad  \mathbf{b}=\left (\begin{array}{c} b_1 \\ b_2  \end{array} \right ) \,.
\end{equation}
The interest in system (\ref{2vecteq}), or rather in its reduced version (see also Section \ref{sec:MI_CNLS}), is motivated by both its wide applicability and by the fact that its NLS one-component version, for $u_2=v_2=0, v_1=- u_1$, turns out to be a good model of the Benjamin-Feir (or modulational) instability which is of great physical relevance \cite{BF1967,HO1972}.
This kind of instability of the plane wave solution of the NLS equation occurs only in the self-focusing regime. In contrast, in the coupled NLS equations the cross interaction and the counterpropagation ($q\neq 0$) introduce additional features (\textit{e.g.}, see  \cite{FMMW2000,BCDLOW2014}) which have no analogues in the NLS equation. This is  so if the instability occurs even when the self-interaction terms have defocusing effects (see Section \ref{sec:MI_CNLS}).
Moreover, and by comparing the coupled case with the single field as in the  NLS equation,  we observe that this CW solution (\ref{2vectsolut}) depends on the  real amplitudes $a_1$,  $a_2$,  $b_1$, $b_2$, and, in a crucial manner (see below), on the real parameter $q$ which measures the \emph{wave-number mismatch} between the two wave components $u_1$ and $u_2$ (or $v_1$ and $v_2$).

The main focus of this section is understanding how the spectrum $\textbf{S}_x$ changes by varying the parameters $a_1$, $a_2$, $b_1$, $b_2$ and $q$.  The results we obtain here will be specialized to the CNLS system in Section \ref{sec:MI_CNLS}.
In matrix notation, see (\ref{matrixQ}), this continuous wave solution (\ref{2vectsolut}) reads
\begin{equation}
\label{cwQ}
Q=R \,\Xi \,R^{-1}\,,\quad \Xi =\left (\begin{array}{ccc} 0 & b_1 & b_2 \\ a_1 & 0 & 0 \\ a_2 & 0 & 0 \end{array} \right )\,,\quad R(x,t) = e^{i(qx \sigma-q^2t \sigma^2 + pt \Sigma)}\,,
\end{equation}
where the matrix $\Sigma$ has expression (\ref{matrixsigma}) while the matrix $\sigma$ is
\begin{equation}\label{matrixsig}
\sigma=\textrm{diag}\{0 \,,\, 1\,,\, -1 \}
\end{equation}
and we conveniently introduce the real parameters
\begin{subequations}
\label{eq:parameters}
\begin{equation}
\label{parap}
p= b_1a_1+b_2 a_2\,
\end{equation}
\begin{equation}\label{parar}
r= b_1a_1-b_2 a_2\,
\end{equation}
\end{subequations}
which will be handy in the following.
Next we observe that a fundamental matrix solution $\Psi(x,t,\lambda)$ of the Lax equations has the expression
\begin{equation}\label{cwPsi}
\Psi(x,t,\lambda)=R(x,t) e^{i(xW(\lambda)-t Z(\lambda))}\,,
\end{equation}
 where the $x$, $t$-independent matrices $W$ and $Z$ are found to be
 \begin{equation}\label{Wmat}
W(\lambda)=\left (\begin{array}{ccc} \lambda & -ib_1 & -i b_2 \\ -i a_1 & -\lambda -q & 0 \\ -ia_2 & 0 & -\lambda +q \end{array} \right ) = \lambda \Sigma -q \sigma -i \,\Xi   \,,
\end{equation}
\begin{equation}
\label{Zmat}
Z(\lambda)=\left (\begin{array}{ccc} -2\lambda^2  & i(2\lambda - q)b_1 & i(2\lambda +q)b_2 \\ i(2\lambda -q) a_1 & 2\lambda^2 -q^2 -a_2 b_2 &  a_1b_2 \\ i(2\lambda+q) a_2 &  a_2 b_1 & 2\lambda^2 -q^2 - a_1 b_1 \end{array} \right )=\lambda^2- 2\lambda W(\lambda) -W^2(\lambda) -p \,,
\end{equation}
with the property that they commute, $[W\,,\,Z]=0$, consistently with the compatibility condition (\ref{compat}). In order to proceed with plain arguments, we consider here the eigenvalues $w_j(\lambda)$ and $z_j(\lambda)$, $j=1,2,3$, of $W(\lambda)$ and, respectively, of $Z(\lambda)$ as simple, as indeed they are for generic values of $\lambda$. In this case, both $W(\lambda)$ and $Z(\lambda)$ are diagonalized by the same matrix $U(\lambda)$, namely
\begin{align}
\label{diagWZ}
&W (\lambda)=U(\lambda)W_D(\lambda) U^{-1}(\lambda)\,,\quad W_D=\textrm{diag}\{w_1,\,w_2 ,\,w_3\} \nonumber\\
&Z(\lambda)=U(\lambda)Z_D(\lambda) U^{-1}(\lambda)\,,\quad Z_D=\textrm{diag}\{z_1,\,z_2 ,\,z_3\} \,.
\end{align}
Next we  construct the matrix $F(x, t, \lambda)$  via its definition, see (\ref{Fmatrix}) and (\ref{phi}),
\begin{equation}\label{Fmatr}
F(x, t, \lambda)=[\Sigma\,,\, \Psi(x,t,\lambda) M(\lambda) \Psi^{-1}(x,t,\lambda)] \,,
\end{equation}
which, because of the explicit expression (\ref{cwPsi}), reads
\begin{equation}\label{FMATR}
F(x, t, \lambda)=R(x,t) \left [\Sigma\,,\,  e^{i(xW(\lambda)-t Z(\lambda))} M(\lambda) e^{-i(xW(\lambda)-t Z(\lambda))} \right ] R^{-1}(x,t) \,.
\end{equation}
As for the matrix $M(\lambda)$, it lies in  a nine-dimensional linear space whose standard basis is given by the matrices $B^{(jm)}$, whose entries are
\begin{equation}\label{basis}
B^{(jm)}_{kn}= \delta_{jk} \delta_{mn}\,,\quad j,\,k,\,m,\,n = 1,\,2,\,3,
\end{equation}
where $\delta_{jk}$ is the Kronecker symbol ($\delta_{jk}=1$ if $j=k$ and $\delta_{jk}=0$ otherwise). However, the  alternative basis  $V^{(jm)}$, which is obtained via the similarity transformation
\begin{equation}\label{newbasis}
V^{(jm)}(\lambda)= U(\lambda) B^{(jm)} U^{-1}(\lambda) \,,
\end{equation}
where $U(\lambda)$ diagonalizes $W$ and $Z$ (see (\ref{diagWZ})),
is more convenient to our purpose. Indeed, expanding the generic matrix $M(\lambda)$ in this basis as
\begin{equation}\label{Mexpan}
M(\lambda)= \sum_{j,m=1}^3 \mu_{jm}(\lambda) V^{(jm)}(\lambda)\,,
\end{equation}
the scalar functions $\mu_{jm}$ being its components, and
inserting this decomposition into expression (\ref{FMATR}), leads to the following representation of $F$
\begin{equation}
\label{Fexpan}
F(x,t,\lambda)=R(x,t) \sum_{j,m=1}^3 \mu_{jm}(\lambda) e^{i[(x(w_j-w_m)-t(z_j-z_m)]} F^{(jm)}(\lambda)R^{-1}(x,t) \,,
\end{equation}
where we have introduced the $x$, $t$-independent matrices
\begin{equation}\label{Fjm}
F^{(jm)}(\lambda)=\left [\Sigma\,,\, V^{(jm)}(\lambda) \right ]\,.
\end{equation}
The advantage of expression (\ref{Fexpan}) is to explicitly show  the dependence of the matrix $F$ on the six exponentials $e^{i[(x(w_j-w_m)-t(z_j-z_m)]}$.

Proposition \ref{prop:4} stated in the previous section guarantees that, for any choice of the functions $ \mu_{jm}(\lambda)$, the expression (\ref{Fexpan}) be a solution of the linearized equation (\ref{linQeq}), see (\ref{Feq}).
It is plain (see (\ref{lincomb})) that the requirement that such solution $\delta Q(x,t)$ be localized in the variable $x$ implies the necessary condition that the functions $\mu_{jm}(\lambda)$ be vanishing for $j=m$, $\mu_{jj}=0$, $j=1, 2, 3$.
The further condition that the solution $\delta Q(x,t)$ be bounded in $x$ at any fixed time $t$ results
in integrating expression (\ref{Fexpan}) with respect to the variable $\lambda$ over the spectral curve  $\textbf{S}_x$  of the complex $\lambda$-plane (see (\ref{lincomb})):
\begin{equation}
\label{lincomb_Sx}
\delta Q(x,t)=\int_{\textbf{S}_x} \mathrm{d}\lambda \,F(x,t,\lambda)\,.
\end{equation}
Here, according to Section \ref{subsec:spectrum}, $\textbf{S}_x$ can be geometrically defined as follows:
\begin{Definition}
\label{def:spectrum}
The $x$-spectrum $\textbf{S}_x$, namely the spectral curve on the complex $\lambda$-plane,  is the set of values of the spectral variable $\lambda$ such that at least one of the three complex numbers $k_j=w_{j+1}- w_{j+2}$, $j=1, 2, 3\, (\textrm{mod}\, 3)$, or explicitly
\begin{equation}\label{k}
k_1(\lambda)=w_2(\lambda) - w_3(\lambda)\,,\quad
k_2(\lambda)=w_3(\lambda) - w_1(\lambda)\,,\quad
k_3(\lambda)=w_1(\lambda) - w_2(\lambda)\;,
\end{equation}
is real.
\end{Definition}
Observe that the $k_j$'s play the role of eigenmode wave-numbers (see (\ref{Fexpan})).

To the purpose of establishing the stability properties of the continuous wave solution (\ref{2vectsolut}) we do not need to compute the integral representation (\ref{lincomb_Sx}) of the solution  $\delta Q$ of (\ref{linQeq}). Indeed, it is sufficient to compute the eigenfrequecies
\begin{equation}
\label{omega}
\omega_1(\lambda) =z_2(\lambda) - z_3(\lambda)\,,\quad
\omega_2(\lambda)=z_3(\lambda) - z_1(\lambda)\,,\quad
\omega_3(\lambda)= z_1(\lambda) - z_2(\lambda)\,,
\end{equation}
as suggested by the exponentials which appear in (\ref{Fexpan}). Their expression follows from the matrix relation (\ref{Zmat})
\begin{equation}\label{zroots}
z_j=\lambda^2 -2\lambda w_j -w^2_j-p\,,
\end{equation}
 and read
\begin{equation}\label{eigenfreq}
\omega_j=-k_j(2\lambda+w_{j+1} +w_{j+2})\,,\quad j=1, 2, 3\, (\textrm{mod}\, 3)\,.
\end{equation}
This expression looks even simpler by using the relation $w_1+w_2+w_3=-\lambda$ implied by the trace of the matrix $W(\lambda)$ (see (\ref{Wmat})) and finally reads
\begin{equation}
\label{eigenFREQ}
\omega_j=k_j(w_j - \lambda)\,\,,\;\; j=1,\, 2,\, 3\,\,.
\end{equation}
The consequence of this expression (\ref{eigenFREQ}), which is relevant to our stability analysis, is given by the following

\begin{Proposition}
\label{prop:5}
The continuous wave solution (\ref{2vectsolut}) is \emph{stable} against perturbations $\delta Q$ whose representation (\ref{lincomb}) is given by an integral  which runs \emph{only} over those  values of $\lambda \in \textbf{S}_x$  which are strictly \emph{real}.
\end{Proposition}
\textbf{Proof}\hspace{0.3cm} If the spectral variable $\lambda$ is real, then all coefficients of the characteristic polynomial of the matrix $W(\lambda)$ (\ref{Wmat}),
\begin{equation}
\label{polyW}
P_W(w;\lambda)= \textrm{det}[w  \mathds{1} - W(\lambda)] =w^3+\lambda w^2+(p-q^2-\lambda^2)w -\lambda^3 +(p+q^2) \lambda -qr \,,
\end{equation}
are real. Indeed, these coefficients depend on the real parameters of the CW solution  (\ref{2vectsolut}), namely the wave number mismatch $q$,  and the parameters $p$ and $r$ defined by (\ref{eq:parameters}). Therefore, the roots $w_j$ of $P_W(w;\lambda)$ are either all real, or one real and two complex conjugate. In the first case, the three wave-numbers $k_j$ (\ref{k}) are all real, and thus the corresponding $\lambda$ is in the spectrum $\textbf{S}_x$, $\lambda\in\mathbf{S}_x$. In the second case, the three $k_j$ are all complex, \textit{i.e.} with non-vanishing imaginary part, and thus $\lambda$ lies in a forbidden interval of the real axis which does not belong to the spectrum $\textbf{S}_x$, $\lambda \notin \textbf{S}_x$. In the following, we refer to one such forbidden real interval as a \emph{gap}, see the Subsection \ref{subsec:gaps} below. Consequently, in the first case, since $\lambda$, $w_1$, $w_2$, $w_3$ and therefore, $k_1$, $k_2$, $k_3$, are all real, then also $\omega_1$, $\omega_2$, $\omega_3$, see (\ref{eigenFREQ}), are all real, with the implication of stability. Indeed, all eigenmode matrices (\ref{Fexpan}) remain small at all times if they are so at the initial time. \\
On the other hand, let us assume now that $\lambda$ is complex, $\lambda=\mu+i\,\rho$, with non-vanishing imaginary part, $\rho\neq0$. Let $w_{j}=\alpha_{j}+i\,\beta_{j}$ be the (generically complex) roots of $P_W(w;\lambda)$. With this notation, we have that one of the wave-numbers, say $k_{3}$, will be real only if $\beta_{1}=\beta_{2}=\beta$. Then, from (\ref{eigenFREQ}), we have that $\omega_{3}$ will also be real only if $\beta_{3}=\rho$. Writing the polynomial $P_W(w;\lambda)$ as $P_W(w;\lambda)=\prod_{j=1}^{3}(w-\alpha_{j}-i\,\beta_{j})$, and comparing the real and imaginary parts of the coefficients of same powers of $w$ from this expression with those obtained from (\ref{polyW}), we get a system of six polynomial equations for the six unknowns $\alpha_{1}$, $\alpha_{2}$, $\alpha_{3}$, $\beta$, $\mu$, and $\rho$, each equation being homogeneous and of degree 1, 2 or 3 in the unknowns:
\begin{align*}
&\beta\,\rho\,(\alpha_1 + \alpha_2) + \alpha_3\,\beta^2 -\alpha_1\,\alpha_2\,\alpha_3 = \mu\,\left(p + q^{2} - \mu^2 + 3\,\rho^2\right)-r\,,\\
&\rho\,\left(\beta^2-\alpha_1\,\alpha_2\right)-\beta\,\alpha_3\,\left(\alpha_1 + \alpha_2\right) = \rho\,\left(p + q^2 - 3\,\mu^2 + \rho^2\right)\,,\\
&\alpha_1\,\alpha_2 + \alpha_2\,\alpha_3 + \alpha_3\,\alpha_1 - \beta\,(\beta + 2\,\rho) = p - q^{2} - \mu^2 + \rho^2\,,\\
&\beta\,\left(\alpha_1 + \alpha_2 + 2\,\alpha_3\right) + \rho\,\left(\alpha_1 + \alpha_2\right) = -2\,\mu\,\rho\,,\\
&-\left(\alpha_1 + \alpha_2 + \alpha_3\right) = \mu\,,\\
& -\left(2 \beta + \rho\right) = \rho\,.
\end{align*}
Then, it is immediate to show by means of elementary algebraic manipulations, that the above system has real solutions for $\alpha_{1}$, $\alpha_{2}$, $\alpha_{3}$, $\beta$, $\mu$, and $\rho$, only if $\rho=-\beta=0$ (unless the non-physical and non-generic condition $p=r=0$ be met). This contradicts the original assumption $\rho\neq0$.
Therefore, for a representation (\ref{lincomb}) of a perturbation $\delta Q$ to be bounded (and thus for the corresponding CW solution to be stable), the integral in (\ref{lincomb}) must run only on those values of $\lambda \in \textbf{S}_x$  which are strictly \emph{real}. \hfill$\square$

This Proposition \ref{prop:5} implies that a real part of the spectrum $\textbf{S}_x$ is always present, and this part may or may not have gaps (see the next Subsection \ref{subsec:gaps}). On the contrary, as it will be proved (see Subsection \ref{subsec:branchesandloops}), a complex component of the spectrum, namely one which lies off the real axis of the $\lambda$-plane, may occur and it always leads to instabilities. This important part of the spectrum $\textbf{S}_x$ is made of open curves, which will be referred to as \emph{branches}, and/or closed curves of the complex $\lambda$-plane which will be termed \emph{loops}.

Before proceeding further we observe that the effect of the cross interaction is crucially influent on the stability only if the phases of the two continuous waves (\ref{2vectsolut}) have different $x$-dependence, namely if $q\neq 0$ (see (\ref{2vectsolut})). Indeed, if $q=0$ the stability property of the continuous wave is essentially that of the NLS equation. This conclusion results from the explicit formulae $w_1= \sqrt{\lambda^2-p}$, $\,w_2= -\sqrt{\lambda^2-p}$, $\,w_3= -\lambda$, and therefore
\begin{equation}
\label{q0kz}
\begin{array}{lll}
\hspace{-0.1cm}k_1=\lambda - \sqrt{\lambda^2-p}\,, & k_2=-\lambda - \sqrt{\lambda^2-p}\,,& k_3=2 \sqrt{\lambda^2-p}\,, \\
\\
\hspace{-0.1cm}\omega_1 = -2\lambda^2 + p +2\lambda \sqrt{\lambda^2-p}\,,&\omega_2= 2\lambda^2 - p +2\lambda \sqrt{\lambda^2-p}\,,& \omega_3= - 4 \lambda \sqrt{\lambda^2-p}\,,
\end{array}
\end{equation}
which show that if $p>0$ (see (\ref{parap})) the spectrum $\textbf{S}_x$ is the real axis with the exclusion of the gap $\{-\sqrt{p} < \lambda < \sqrt{p}\,\}$, while, if $p<0$, the spectrum $\textbf{S}_x$ is the entire real axis with the addition of the imaginary branch $\lambda=i \rho$, $\{- \sqrt{-p} < \rho< \sqrt{-p}\,\}$. Therefore, the stability for $p>0$ follows from the reality of the frequencies $\omega_1$,  $\omega_2$, $\omega_3$ over the whole spectrum $\textbf{S}_x$, see (\ref{q0kz}).  The modulational instability occurs only if $p<0$ since, for $\lambda$ in the imaginary branch of  the spectrum $\textbf{S}_x$,  the frequencies $\omega_1$, $\omega_2$ and $\omega_3$
have a non vanishing imaginary part.

In the following, while computing the spectrum  $\textbf{S}_x$, we consider therefore only the case $q\neq 0$ and, with no loss of generality, strictly positive, $q>0$. In this respect, we also note that the assumption that $q$ be non-vanishing makes it possible, without any loss of generality, to rescale it to unit, $q=1$, while keeping in mind that, by doing so, $\lambda$, $w_j$ rescale as $q$, while $p$, $r$, $z_j$ rescale as $q^2$, \textit{i.e.}
\[
\lambda\mapsto q\lambda\,,\quad w\mapsto q\,w\,,\quad p\mapsto q^{2}\,p\,,\quad r\mapsto q^{2}\,r\,,\quad z\mapsto q^{2}\,z\,.
\]
However, we deem it helpful to the reader to maintain the parameter $q$ in our formulae, here and in the following, to have a good control of the limit as $q$
vanishes. In addition, here and thereafter, our computations and results will be formulated for non-negative values of the parameter $r$ according to the following
\begin{Proposition}
\label{prop:6}
With no loss of generality, the relevant parameter space reduces to the half $\,(r\,,\,p)$-plane with $r \geq 0$.
\end{Proposition}
Indeed, the change of sign $\lambda \mapsto-\lambda$, $r \mapsto -r$ takes the characteristic polynomial $P_W(w;\lambda)$ (\ref{polyW}) into $-P_W(-w;\lambda)$ and this implies that our attention may be confined to non-negative values of $r$ only.

\subsection{Gaps}
\label{subsec:gaps}
Although computing the position of the gaps of $\textbf{S}_x$ on the real $\lambda$-axis is not strictly relevant to the issue of stability because of Proposition \ref{prop:5}, this information is essential to the construction of soliton solutions corresponding to real discrete eigenvalues which lie inside a gap. With this motivation in mind, we devote this subsection to this task.
Using the Cardano formulae for finding workable explicit expressions of the three roots $w_j(\lambda)$ of $P_W(w;\lambda)$ for each value of $p$, $r$ and $q$ is not practical. To overcome this difficulty, we will adopt here an algebraic-geometric approach.

We begin by computing the discriminant $D_W=\Delta_{w}P_W(w;\lambda)$ of the polynomial $P_W(w;\lambda)$ (\ref{polyW}) with respect to $w$, obtaining thus a polynomial in $\lambda$, with parameters $q$, $p$, $r$,
\begin{equation}\label{discrW}
D_W(\lambda;q,p,r)= 64 q^2 \lambda^4 -32 q r \lambda^3 + 4(p^2-20 q^2 p-8 q^4) \lambda^2 +36 q r (2 q^2+p) \lambda -4(p-q^2)^3 -27 q^2 r^2\,,
\end{equation}
which is \emph{positive} whenever the three roots $w_j$ are real, and it is \emph{negative} if instead only one root is real. Consequently, for any given value of the parameters $q$, $p$, and $r$, the discriminant is negative, $D_W<0$, for those real values of $\lambda$ which belong to a gap of $\textbf{S}_x$, while its zeros are the end points of gaps. Here and hereafter, the notation $\Delta_{y}P(x,y,z)$ stands for the discriminant of the polynomial $P$ with respect to its variable $y$.

In the exceptional case $r=0$, computing the roots of the discriminant $D_W$ as a polynomial in the variable $\lambda$ reduces to the factorization of a quadratic polynomial, allowing the formulation of the following
\begin{Proposition}
\label{prop:7} For $r=0$, the gaps depend on the parameter $p$ as follows:
\begin{center}
\begin{tabular}{rl}
\hline
$-\infty < p < 0$ & no gap\\
\hline
$0 \leq p < q^2$ & two gaps, $-g_+ \leq  \lambda \leq -g_- $ and $g_- \leq  \lambda \leq g_+$\\
\hline
$q^2 \leq p < +\infty$ & one gap, $-g_+ \leq \lambda \leq g_+$\\
\hline
\end{tabular}
\end{center}
The border values $g_{\pm}$ are
\begin{equation}
\label{gap}
g_{\pm}= \frac{\sqrt{2}}{8 q} \sqrt{8 q^4+20 q^2 p -p^2 \pm \sqrt{p\,(p+8 q^2)^3}}\,\,.
\end{equation}
The threshold values $p=0$ and $p=q^2$ correspond to the opening and the coalescence of two gaps, respectively.
\end{Proposition}
In the generic case $r\geq0$, we find that the number of gaps of the real part of the  spectrum $\textbf{S}_x$ is either zero, or one, or two. Gaps appear at double-zeros of the discriminant $D_W(\lambda;q,p,r)$, thus at zeros of its own discriminant $\Delta_{\lambda}\,D_W(\lambda;q,p,r)=\Delta_{\lambda}\,\Delta_w P_{W}(w;\lambda)$, namely when
\begin{equation}
\label{eq:disc_disc}
\Delta_{\lambda}\,\,D_{W}(\lambda;q,p,r) =
-1048576\,q^{2}\,(p-r)\,(p + r)\,\left[\left(p-q^{2}\right)\,\left(p+8\,q^{2}\right)^2-27\,q^{2}\,r^{2}\right]^{3}=0\,.
\end{equation}
The three polynomial factors appearing in (\ref{eq:disc_disc}) bound the regions of the $(r,\,p)$-plane characterized by different numbers of gaps.

By varying the values of the parameters $q$, $p$, and $r$, we expect the double-zeros to open and form the gaps, namely intervals where $D_W<0$. In order to find the number of gaps from the expression (\ref{discrW}) of the discriminant $D_W(\lambda;q,p,r)$, it is convenient to plot the dependence of the parameter $p=p(\lambda)$ on the spectral variable $\lambda$  at the zeros of this discriminant for fixed values of the parameters $r$ and $q$. This function $p(\lambda)$ is therefore implicitly defined by the equation $D_W(\lambda;q,p(\lambda),r)=0$.
For a fixed value of $q$, in the $(\lambda,\,p)$-plane the $p(\lambda)$ curve has always one cusp (a double singular point), and either two minima, or one minimum and one maximum, depending on the value of $r$ (see Appendix \ref{app:A} for details). The position of the cusp $( \lambda_S,\, p_S)$ as a function of $r$ and $q$ is
 \begin{equation}
 \label{expli}
 \begin{array}{l}
\lambda_S(r)=\frac32 \left \{\left [(qr/2)+\sqrt{q^6+(qr/2)^2}\right ]^{1/3} - q^2\left [(qr/2)+\sqrt{q^6+(qr/2)^2}\right ]^{-1/3}\right \}\,,\\
 p_S(r)=q^2+\frac{4}{3}\lambda_{S}^2(r)\,.
 \end{array}
 \end{equation}
Furthermore, $p_S(r)$ seen as an algebraic curve in the $(r,\,p)$-plane satisfies the following implicit relation
\begin{equation}
\label{p_S_implicit}
(p-q^2)\,(p+8q^2)^2 -27\,r^2\,q^2=0\,,\quad p=p_S(r)\,.
\end{equation}
The transition between the two regimes (two minima, or one minimum and one maximum) takes place at the special threshold value $r_T=4q^2$ where $p_S(r_T)=r_T$ (see Appendix \ref{app:A}). If $0\leq r<r_T$, the cusp is also a local maximum.
Equipped with these findings, we formulate
\begin{Proposition}
\label{prop:8}
For real $\lambda$ and $r>0$,  $\mathbf{S}_x$ has the following gap structure.

If $r < r_{T}=4q^2$
\begin{center}
\begin{tabular}{ll}
\hline
$- \infty < p < -r $ & no gaps\\
\hline
$-r \leq p < r$   & one gap\\
\hline
$r \leq p < p_S(r)$  & two gaps\\
\hline
$p_S(r) \leq p < +\infty$ & one gap\\
\hline
\end{tabular}
\end{center}
The threshold values $p=-r$, $p=r$, and $p=p_S(r)$  correspond to the opening of one gap, the opening of two gaps, and the coalescence  of two gaps, respectively.

If $r > r_{T}=4q^2$
\begin{center}
\begin{tabular}{ll}
\hline
$- \infty < p < -r $ & no gaps\\
\hline
$-r \leq p < p_S(r)$ & one gap\\
\hline
$p_S(r) \leq p < r$ & two gaps\\
\hline
$r \leq p < +\infty$ & one gap\\
\hline
\end{tabular}
\end{center}
The threshold values $p=-r$, $p = p_S(r)$, and $p=r$ correspond to the opening of one gap, the opening of two gaps, and the coalescence of two gaps, respectively.
\end{Proposition}
\vspace{-0.3cm}
\begin{figure}[h!]
\begin{center}
\includegraphics[width=0.75\textwidth]{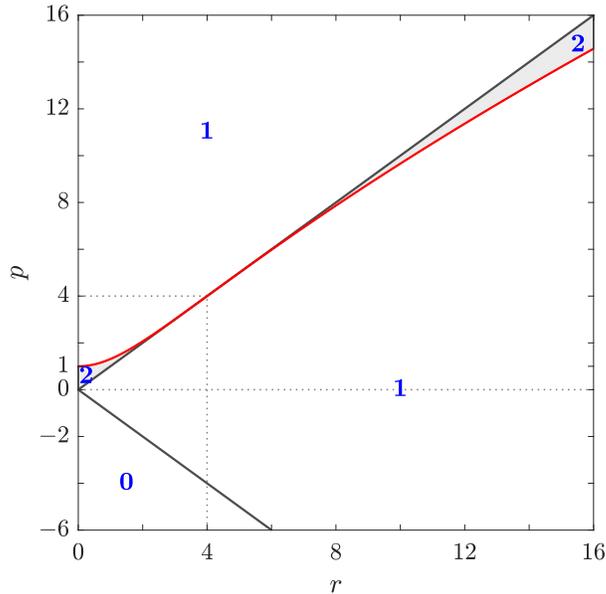}
\caption{The three threshold curves $p_{\pm}(r)=\pm r$ (solid black) and $p_S(r)$ (solid red) for $q=1$, as parametrically defined by (\ref{threshpm}) and (\ref{p_S_implicit}) or explicitly by (\ref{expli}), are plotted. They are boundaries of regions of the $(r,p)$-plane, $r>0$, where the number of gaps, either $0$, or $1$, or $2$, is shown. It is also shown that $ p_S(r)>r$ for $r<4$, and that $ p_S(r)<r$ for $r>4$ (see Proposition \ref{prop:8}).}
\label{fig:pS(r)}
\end{center}
\end{figure}

As implied by (\ref{eq:disc_disc}) and Proposition \ref{prop:8}, we identify three threshold curves in the $(r,p)$-half-plane; they are
\begin{equation}
\label{threshpm}
 p=p_{\pm}(r)=\pm r
 \end{equation}
and the curve $p=p_S(r)$ which is explicitly expressed by (\ref{expli}) and plotted in Figure \ref{fig:pS(r)}. Note that, according to Proposition \ref{prop:6}, we will focus only on the half-plane $r \geq 0$.  These curves will play a role also in the next subsection where we give a complete classification of branches and loops of the spectrum.

\subsection{Branches and loops}
\label{subsec:branchesandloops}
In the previous subsection we have described the gaps of the spectrum $\textbf{S}_x$, namely those values of the spectral parameter $\lambda$ which are real but do not belong to the spectrum. Here we face instead the problem of finding the subset of the $\lambda$-plane which is off the real axis but belongs to $\textbf{S}_x$. This subset is made of smooth and generically finite, open (branches) or closed (loops), continuous curves (with the possible exception of the case $r=0$, see Figure \ref{fig:r=03} and \cite{DLS2017}). By Definition \ref{def:spectrum}, $\lambda \in \textbf{S}_x$ if at least one of the corresponding three wave numbers $k_j(\lambda)$, $j=1, 2, 3$ is real, see (\ref{k}). Going back to the characteristic polynomial $P_W(w;\lambda)$ (\ref{polyW}), we look now at its roots $w_j$, and at the wave numbers $k_j$, $j=1, 2, 3$, see (\ref{k}).
If $\lambda$ is not real, the roots $w_j$ cannot be all real themselves since the coefficients of $P_W(w;\lambda)$ are not real. As a consequence, the requirement that one of the wave numbers (\ref{k}) be real implies that at least two roots of $P_W(w;\lambda)$, for instance $w_1$ and $w_2$,  have the same imaginary part. In fact, branches and loops of $\textbf{S}_x$ may exist only in this case: indeed, if the three roots $w_j$ have all the same imaginary part, it can be shown that no smooth branch or loop exists.
Hence, here and below, we name $k_3=w_1-w_2$ the only real wave number, while $k_1$ and $k_2$ are complex (note that $k_1+k_2+k_3=0$). The complex part of the spectrum $\textbf{S}_x$ is therefore defined as the set of the $\lambda$-plane such that $\mathrm{Im}(k_3(\lambda))=0$.
In order to compute this component of the spectrum, we introduce the novel polynomial $\mathcal{P}(\zeta)$,
\begin{equation}
\label{polyK}
\mathcal{P}(\zeta)= \zeta^3 + d_1 \zeta^2 +  d_2 \zeta +  d_3\,,
\end{equation}
defined by the requirement that its roots are the squares of the wave numbers, $\zeta_j=k_j^2$. It is plain that its coefficients $d_j$ are completely symmetric functions of the roots $w_j$ of the characteristic polynomial $P_W(w;\lambda)$. This is evidently so because $d_1$, $d_2$, $d_3$ are symmetric functions of $\zeta_1=(w_2-w_3)^2$, $\zeta_2=(w_3-w_1)^2$, $\zeta_3=(w_1-w_2)^2$, and therefore of $w_1$, $w_2$, $w_3$, with the implication that the coefficients $d_1$, $d_2$, $d_3$ of $\mathcal{P}(\zeta)$ are polynomial functions of the coefficients $c_1$, $c_2$, $c_3$ of the characteristic polynomial (\ref{polyW}),
$
P_W(w)= w^3 + c_1 w^2 +  c_2 w +  c_3\,,
$
\begin{equation}
\label{coeffW}
c_1=\lambda \,,\quad c_2 = p-q^2-\lambda^2\,,\quad c_3= -\lambda^3+(p+q^2)\lambda-qr \,.
\end{equation}
These relations, which read
\begin{equation}
\label{c/d}
\begin{array}{lll} d_1&=&2(3c_2-c_1^2)\,,   \\
d_2&=&(3c_2-c_1^2)^2 = \frac14 d_1^2  \,,  \\
d_3&=&(4c_2-c_1^2) (c_2^2-4c_1c_3) +c_3(27c_3-2c_1c_2) \,,
\end{array}
\end{equation}
combined with the expressions of the coefficients $c_j$, see (\ref{coeffW}), yield the dependence of the coefficients $d_j$ of $\mathcal{P}(\zeta)$ on the parameters $r$, $p$, $q$ and on the spectral variable $\lambda$, so that $\mathcal{P}(\zeta)\equiv \mathcal{P}(\zeta;\lambda,q,p,r)$.
This dependence turns out to be
\begin{equation}
\label{dcoeff}
d_1=2[-4\lambda^2 +3(p-q^2)]\,,\quad d_2= [-4\lambda^2 +3(p-q^2)]^2\,,\quad d_3=-D_W(\lambda;q,p,r)\,,
\end{equation}
where $D_W$ is the discriminant of $P_W$, see its expression (\ref{discrW}), with the implication that $\mathcal{P}(\zeta;\lambda,q,p,r)$ is of degree $4$ in the spectral variable $\lambda$.
As explained at the beginning of this section, we recall that, for an arbitrary complex $\lambda$, at most one wave number (named $k_{3}$) is real. Therefore our task of computing the complex part of the spectrum $\textbf{S}_x$, for a given value of the parameters $r$, $p$, reduces to finding the curve in the $\lambda$-plane along which the root $\zeta_3(\lambda) = k_{3}^{2}(\lambda)$ of $\mathcal{P}(\zeta;\lambda,q,p,r)$ remains \emph{real and positive}, $\zeta_3=k_{3}^{2}>0$. However, it is in general much more convenient, from both the analytical and computational points of view, (and indeed this is what we will be doing in the following), to reverse the perspective and regard the $x$-spectrum $\mathbf{S}_{x}$ as the \emph{locus} in the $\lambda$-plane of the $\lambda$-roots of $\mathcal{P}(\zeta;\lambda,q,p,r)$ seen as a \emph{real} polynomial in $\lambda$, for $\zeta$ spanning over the semi-line $[0,\,+\infty)$. In other words, for a fixed value of $\zeta\geq0$, we compute the four $\lambda$-roots of the real polynomial $\mathcal{P}(\zeta;\lambda,q,p,r)$, which are the values of $\lambda$ (irrespective of being complex or purely real) such that the wave number $k_{3}=\sqrt{\zeta}$ is real. In this way, the two cases of $\mathrm{Im}(\lambda)=0$ and $\mathrm{Im}(\lambda)\neq0$ can be treated at once (allowing to retrieve and confirm all the results about gaps in the spectra presented in the previous subsection).

From the algebraic-geometric point of view, the locus of the roots of $\mathcal{P}(\zeta;\lambda,q,p,r)$ in the $\lambda$-plane is an algebraic curve; this can be given implicitly as a system of two polynomial equations in two unknowns by setting $\lambda=\mu+i\,\rho$ and then separating the real and the imaginary parts of $\mathcal{P}(\zeta;\lambda,q,p,r)$.

In order to explain how the spectrum changes over the parameter space, we apply Sturm's chains (\textit{e.g.}, see \cite{demidovich1981computational,HM1990}) to the polynomial $\mathcal{Q}(\zeta;q,p,r)=\Delta_{\lambda}\,\mathcal{P}(\zeta;\lambda,q,p,r)$, that is, to the discriminant of the polynomial $\mathcal{P}(\zeta;\lambda,q,p,r)$ with respect to $\lambda$. Indeed, it turns out that the nature and the number of the components of the $x$-spectrum $\mathbf{S}_{x}$, seen as a curve in the $\lambda$-plane, are classified by the number of sign-changes of $\mathcal{Q}(\zeta;q,p,r)$, for $\zeta\geq 0$. This procedure requires some technical digression, thus, in order to avoid breaking the narrative here, we refer the reader to Appendix \ref{app:B} for all details.

Our  findings provide a complete classification of the spectra over the entire parameter space, \textit{i.e} over the $(r, p)$-plane, $r \geq 0$.
First we note that the number of branches (B) can only be 0, 1, or 2, while there may occur either 0, or 1 loop (L). Moreover, in addition to the three threshold curves $p_{\pm}(r)$ (\ref{threshpm}) and $p_S(r)$ (\ref{expli}) introduced in Subsection \ref{subsec:gaps},  one more threshold curve, $p=p_C(r)$, is found, which, for a given non vanishing value of $q$, is implicitly defined as
\begin{equation}
\label{threshC}
(p^2-16q^4)^3+432q^4r^2(p^2-r^2)=0 \,,\quad p=p_C(r)\,,\quad \text{with}\; p<0\,,
\end{equation}
or, explicitly, as
\begin{equation}
\label{epliC}
p_C(r)=-\sqrt{16q^4-12q^2\,(2qr)^{2/3}+3\,(2qr)^{4/3}}\,.
\end{equation}
It is now convenient to combine these results with those we obtained in the previous subsection on the gaps (G) on the real $\lambda$-axis of the complex $\lambda$-plane, obtaining a complete classification of the spectra.
We find that only five different types of spectra exist according to different combinations of gaps, branches and loops.
These are the following:
\[
\text{0G\, 2B\, 0L\,, \quad 0G\, 2B\, 1L\,,\quad 1G\, 1B\, 0L\,,\quad 1G\, 1B\, 1L\,,\quad 2G\, 0B\, 1L\,,}
\]
where the notation $n$X stands for $n$ components of the type X, with X either G, or B, or L.
\begin{figure}[h!]
\begin{subfigure}[h]{0.5\textwidth}
\hspace{-0.3cm}
\includegraphics[width=1.1\textwidth]{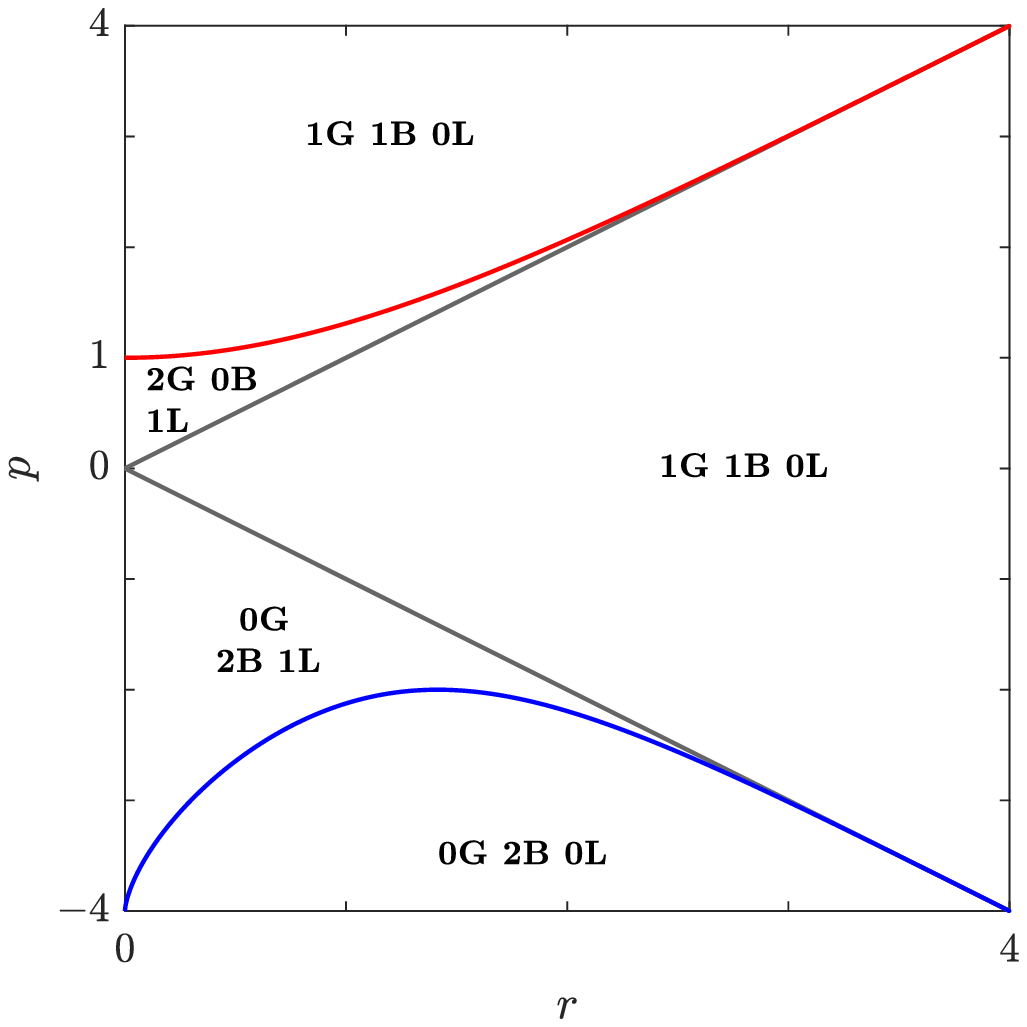}
\caption{$r\leq 4q^2$}
\end{subfigure}
\begin{subfigure}[h]{0.5\textwidth}
\hspace{-0.3cm}
\includegraphics[width=1.1\textwidth]{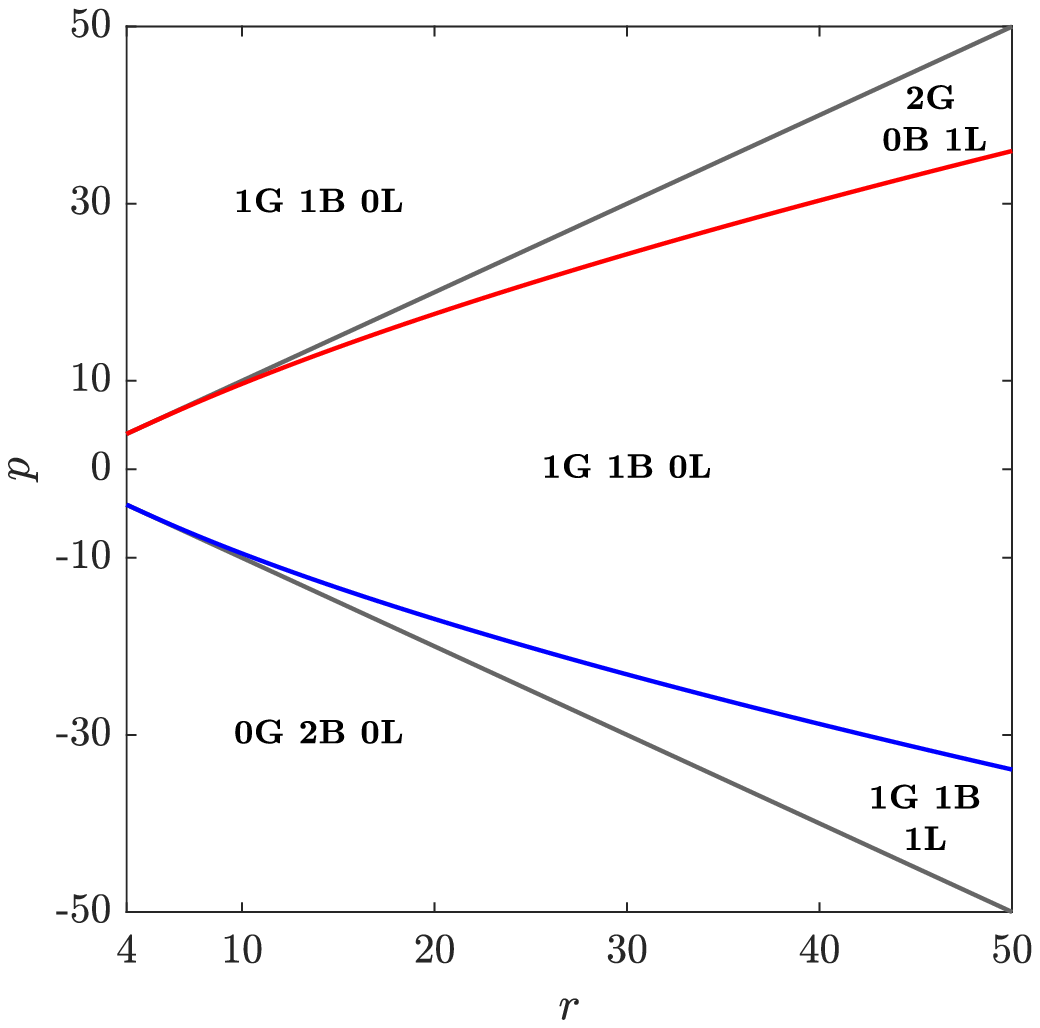}
\caption{$r\geq  4q^2$}
\end{subfigure}
\caption{The structure of the spectrum $\mathbf{S}_x $ in the $(r,p)$ plane, $q=1$ (see Proposition \ref{prop:9}). In red, the curve $p_S$; in blue, the curve $p_C$; in grey, the curves $p=\pm r$. \label{fig:spectrum}}
\end{figure}

In Figure \ref{fig:spectrum} the four threshold curves $p_{\pm}$, $p_S$, $p_C$ (with $q=1$) are plotted to show the partition of the parameter half-plane according to the gap, branch and loop components of the spectrum, for both $r<r_T=4\,q^2$ and $r>r_T=4\,q^2$. This overall structure of the spectrum $\textbf{S}_x$ in the parameter space  can be summarized by the following
\begin{Proposition}
\label{prop:9}
For  $r\geq 0$,  the spectrum $\mathbf{S}_x$ has the following structure in the parameter plane $(r,\,p)$.

If $r < r_{T}=4q^2$
\begin{center}
\begin{tabular}{ll}
\hline
$- \infty < p < p_C(r)$ & $\mathrm{0G\,\,  2B\,\,  0L}$\\
\hline
$ p_C(r) < p < -r$   & $\mathrm{0G\,\,  2B\,\,  1L}$\\
\hline
$-r < p < r$  & $\mathrm{1G\,\, 1B\,\, 0L}$\\
\hline
$ r < p < p_S(r)$ & $\mathrm{2G\,\, 0B\,\, 1L}$\\
\hline
$ p_S(r) < p < +\infty$ & $\mathrm{1G\,\, 1B\,\, 0L}$\\
\hline
\end{tabular}
\end{center}
 If $r > r_{T}=4q^2$
\begin{center}
\begin{tabular}{ll}
\hline
$- \infty < p < -r$ & $\mathrm{0G\,\,  2B\,\,  0L}$\\
\hline
$ -r < p < p_C(r)$   & $\mathrm{1G\,\,  1B\,\,  1L}$\\
\hline
$p_C(r) < p < p_S(r)$  & $\mathrm{1G\,\, 1B\,\, 0L}$\\
\hline
$ p_S(r) < p < r$ & $\mathrm{2G\,\, 0B\,\, 1L}$\\
\hline
$ r < p < +\infty$ & $\mathrm{1G\,\, 1B\,\, 0L}$\\
\hline
\end{tabular}
\end{center}
\end{Proposition}
At the threshold values, the dynamics of the transitions between the five types of spectra can be very rich, and phenomena such as the merging of a branch and a loop to form a single branch can be observed; however, in the simplest cases (which are the majority), gaps, branches and loops open up from, or coalesce into points. An important threshold case is $p=r$ (entailing $a_{2}=0$, and as such non-generic): this is the only choice of the parameters $p$, and $r$ for which the spectrum is entirely real (and thus the CW solution is stable), with one gap in the interval $\left(-\frac{2\,\sqrt{r}+q}{2},\frac{2\,\sqrt{r}-q}{2}\right)$.

Examples of different spectra have been computed numerically by calculating the zeros of $\mathcal{P}(\zeta;\lambda,q,p,r)$ as a polynomial in $\lambda$ (see Appendix \ref{app:C} for details), and are displayed in Figure \ref{fig:spectra}. They are representative of the five types, and correspond to various points of the parameter $(r,\,p)$-space as reported in Proposition \ref{prop:9}.

Finally, the limit case $r=0$ of the parameter half-plane $r>0$ deserves a special mention. In fact, four different limit spectra are found for $r=0$, namely in the four intervals  $-\infty<p<-4\,q^2$, $-4\,q^2<p<0$, $0<p<q^2$ and
$q^2<p<+\infty$.
After meticulous analysis (see \cite{DLS2017}) we find that these limit spectra are consistent with those depicted by Proposition \ref{prop:9}. However, they have to be carefully dealt with as, in this limit, two gaps may coalesce in one, or a loop may go through the point at infinity of the $\lambda$-plane thereby covering the whole imaginary axis.
These spectra are displayed in Figures \ref{fig:r=01}, \ref{fig:r=02} and \ref{fig:r=03}.
We should also point out that the condition $r=0$ allows for analytic computations, and explicit formulae, since, as already noted in Subsection \ref{subsec:gaps} for the discriminant $D_W(\lambda;q,p,r)$, the coefficients $d_j(\lambda)$ of the relevant polynomial $\mathcal{P}(\zeta;\lambda,q,p,r)$ (\ref{polyK}) reduce to polynomials of degree 2 in the variable $\lambda^2$.
\begin{figure}[h!]
\begin{subfigure}[ht]{0.45\textwidth}
\centering
\includegraphics[width=0.95\textwidth]{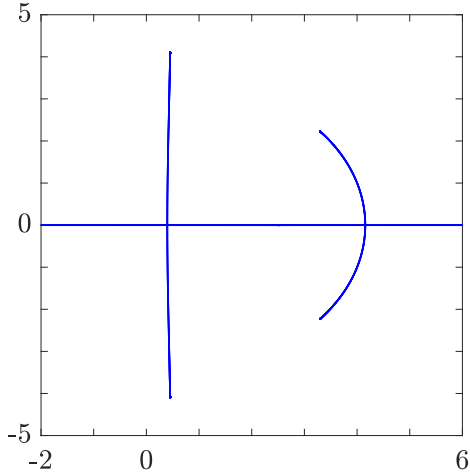}
\caption{$\textbf{S}_x$ for $r=15$, $p=-17$, \textit{i.e.}, $s_{1}=s_{2}=-1$, $a_{1}=1$, $a_{2}=4$, as an example of a 0G 2B 0L spectrum.\vspace{0.5cm}}
\end{subfigure}\qquad
\begin{subfigure}[ht]{0.45\textwidth}
\centering
\includegraphics[width=0.99\textwidth]{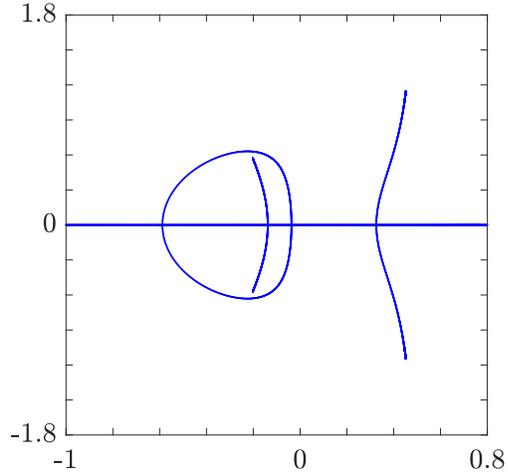}
\caption{$\textbf{S}_x$ for $r=1$, $p=-1.5$, \textit{i.e.}, $s_{1}=s_{2}=-1$, $a_{1}=0.5$, $a_{2}=1.118$, as an example of a 0G 2B 1L spectrum.\vspace{0.5cm}}
\end{subfigure}\\
\begin{subfigure}[ht]{0.45\textwidth}
\centering
\includegraphics[width=0.95\textwidth]{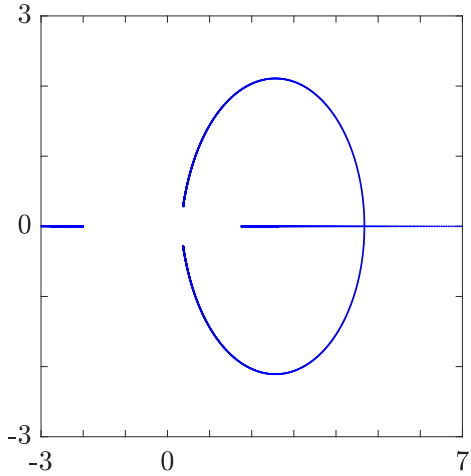}
\caption{$\textbf{S}_x$ for $r=1$, $p=2.8$, \textit{i.e.}, $s_{1}=s_{2}=1$, $a_{1}=1.3784$, $a_{2}=0.94868$, as an example of a 1G 1B 0L spectrum.}
\end{subfigure}\qquad
\begin{subfigure}[ht]{0.45\textwidth}
\centering
\includegraphics[width=0.95\textwidth]{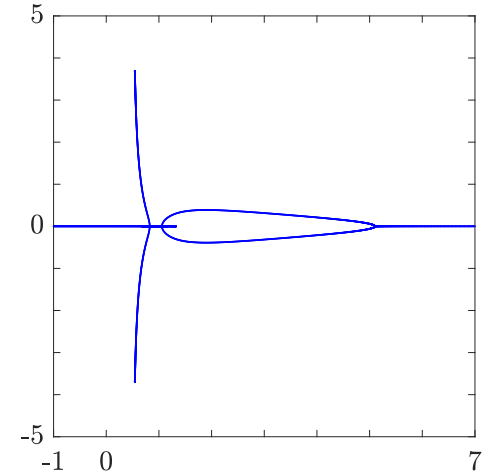}
\caption{$\textbf{S}_x$ for $r=15$, $p=-13.45$, \textit{i.e.}, $s_{1}=1$, $s_{2}=-1$, $a_{1}=0.88034$, $a_{2}=3.7716$, as an example of a 1G 1B 1L spectrum.}
\end{subfigure}
\caption{Examples of spectra for different values of $r$ and $p$, with $q=1$. \label{fig:spectra}}
\end{figure}
\begin{figure}[h!]
\ContinuedFloat
\begin{subfigure}[ht]{0.45\textwidth}
\centering
\includegraphics[width=0.99\textwidth]{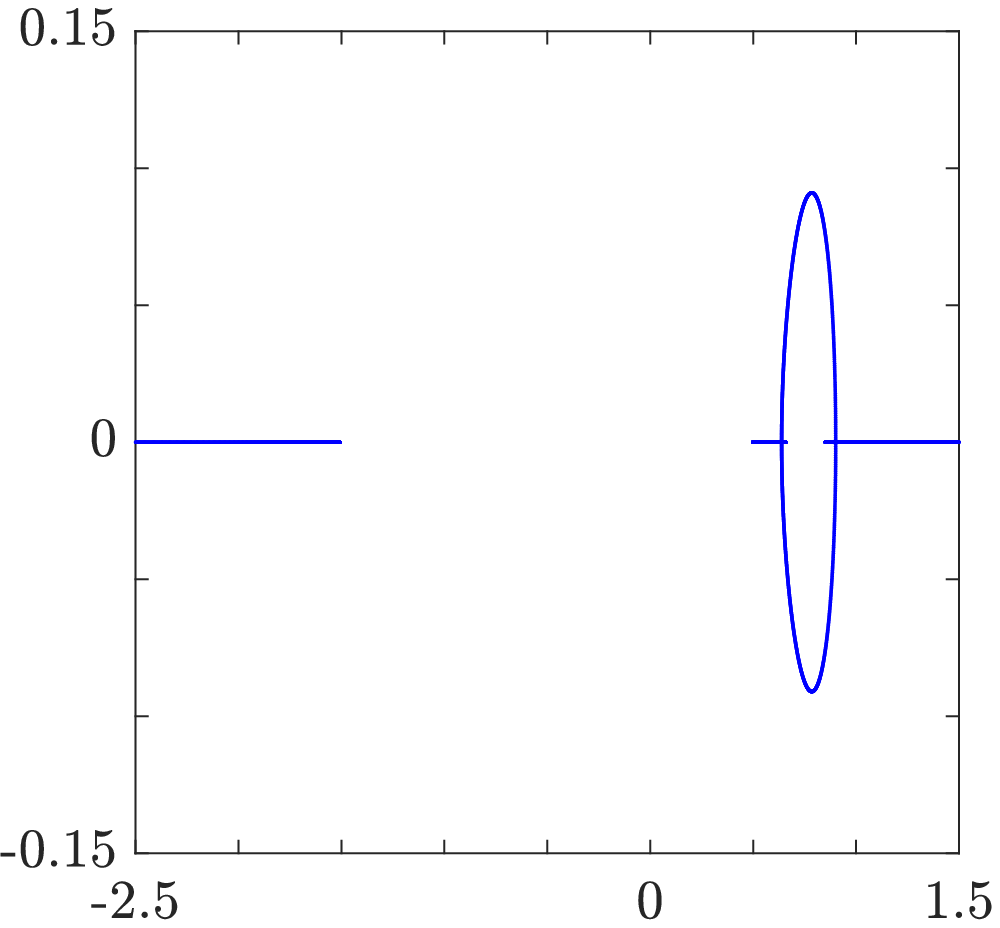}
\caption{$\textbf{S}_x$ for $r=1$, $p=1.025$, \textit{i.e.}, $s_{1}=s_{2}=1$, $a_{1}=1.0062$, $a_{2}=0.1118$, as an example of a 2G 0B 1L spectrum.\\ \\\vspace{0.5cm}}
\end{subfigure}\qquad
\begin{subfigure}[ht]{0.45\textwidth}
\centering
\includegraphics[width=0.95\textwidth]{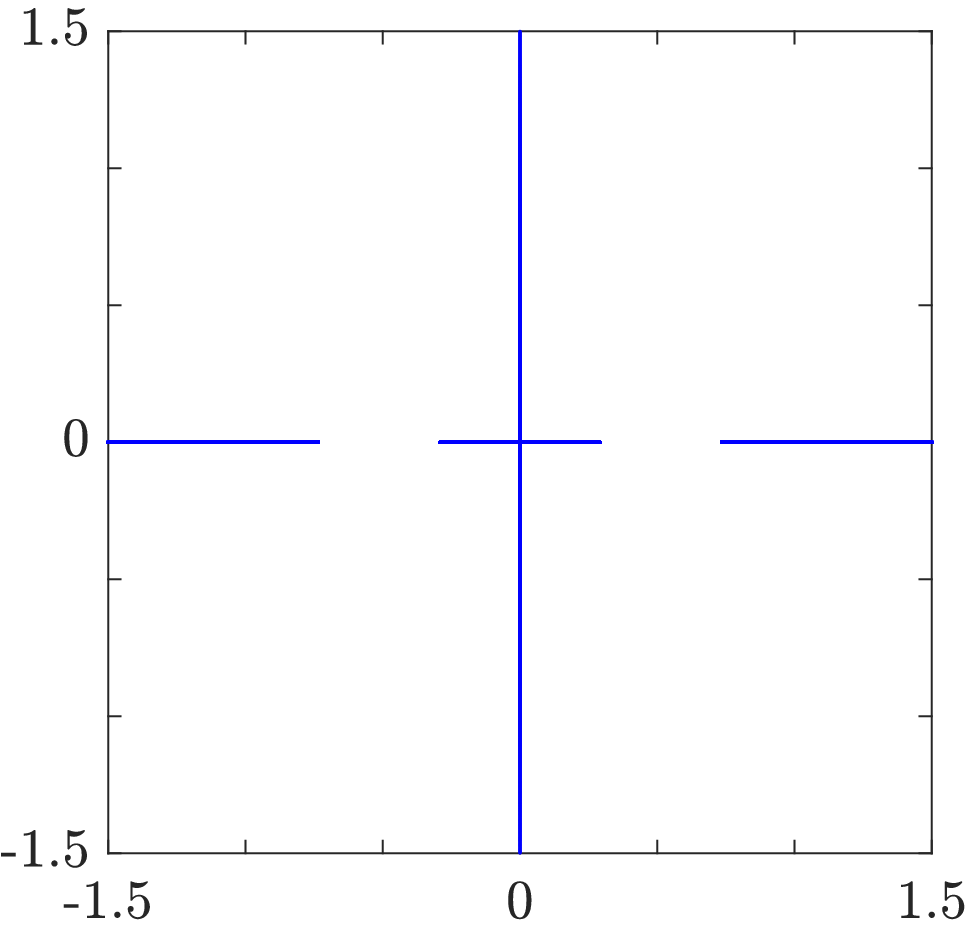}
\caption{$\textbf{S}_x$ for $r=0$, $p=0.1$, \textit{i.e.}, $s_{1}=s_{2}=1$, $a_{1}=a_{2}=0.22361$, as an example of a degenerate case of a 2G 0B 1L spectrum: the imaginary axis in the spectrum is a loop passing through the point at infinity.\vspace{0.5cm}\label{fig:r=01}}
\end{subfigure}\\
\begin{subfigure}[ht]{0.45\textwidth}
\centering
\includegraphics[width=0.95\textwidth]{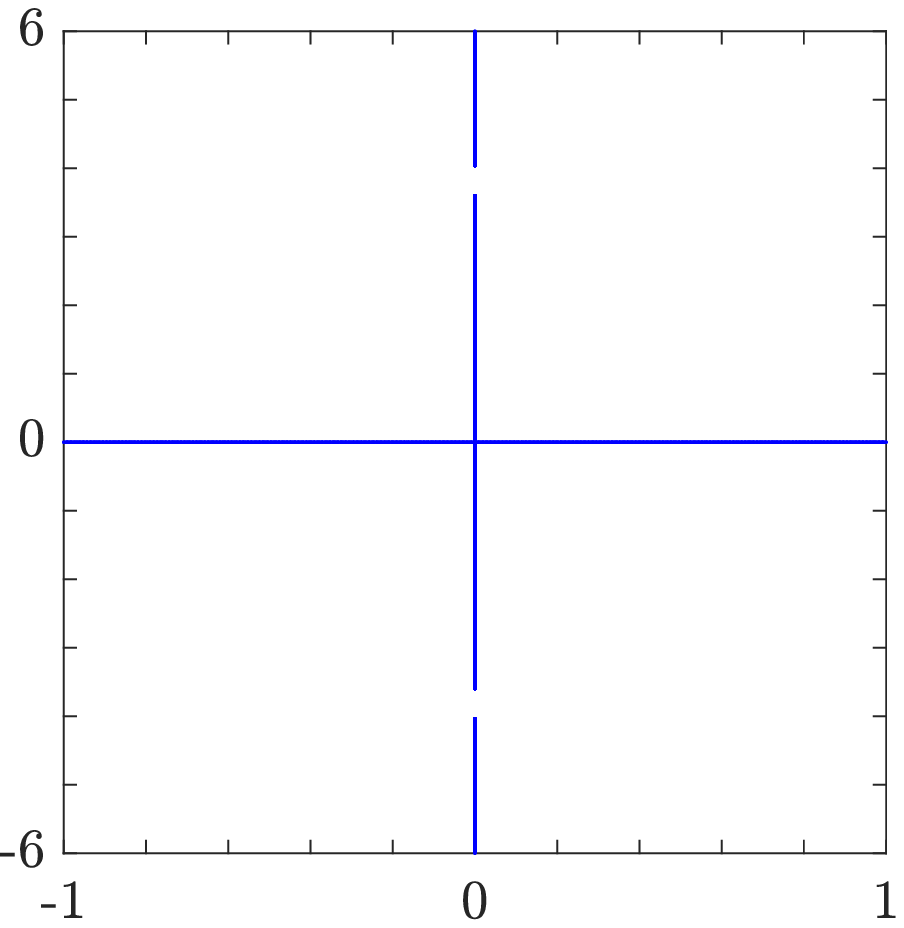}
\caption{$\textbf{S}_x$ for $r=0$, $p=-14$, \textit{i.e.}, $s_{1}=s_{2}=-1$, $a_{1}=a_{2}=2.6458$, as an example of a degenerate case of a 0G 2B 0L spectrum: both branches are entirely contained on the imaginary axis; one branch passes through the point at infinity, whereas the other branch passes through the origin; for $r=0$, two symmetrical gaps open on the imaginary axis for $p\leq-8$, as explained in \cite{DLS2017}.\label{fig:r=02}\\}
\end{subfigure}\qquad
\begin{subfigure}[ht]{0.45\textwidth}
\centering
\includegraphics[width=0.95\textwidth]{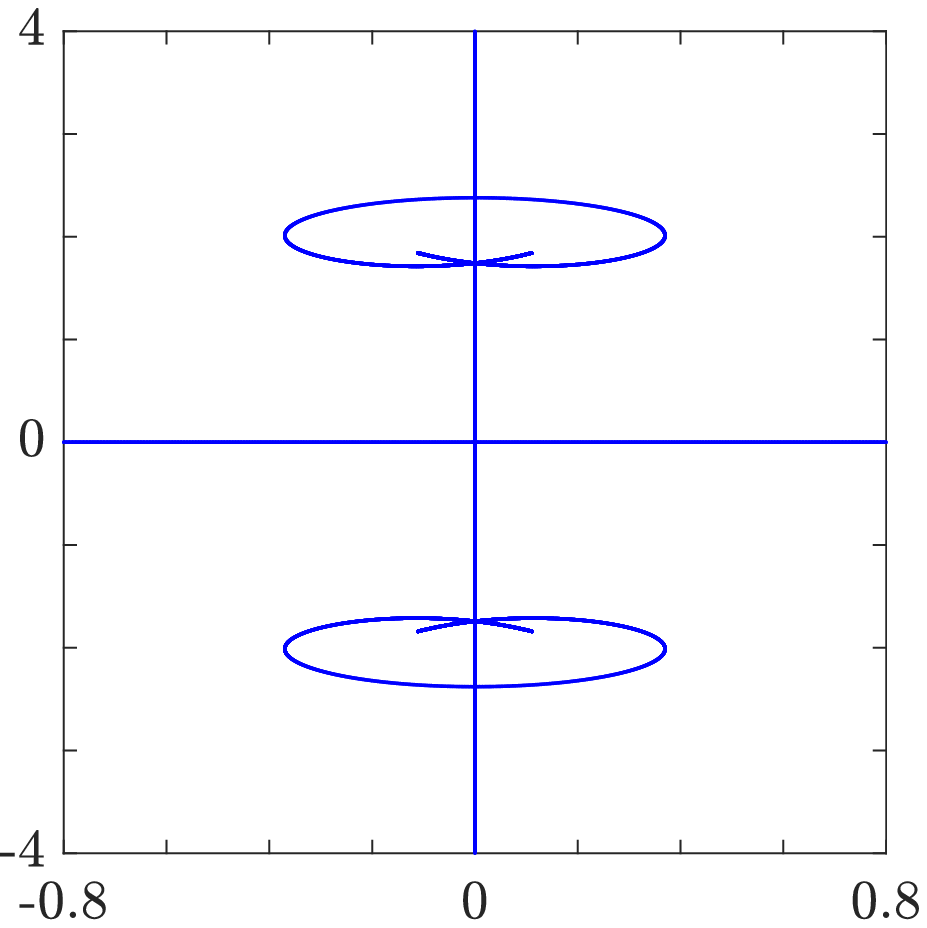}
\caption{$\textbf{S}_x$ for $r=0$, $p=-4.7$, \textit{i.e.}, $s_{1}=s_{2}=-1$, $a_{1}=a_{2}=1.533$, as an example of a degenerate case of a 0G 2B 0L spectrum: this case (which also appears in \cite{LZ2017}) when projected back onto the stereographic sphere, can be completely explained in terms of the classification scheme provided in Proposition \ref{prop:9} (see \cite{DLS2017}).\\ \\ \label{fig:r=03}}
\end{subfigure}
\caption{Examples of spectra for different values of $r$ and $p$, with $q=1$.}
\end{figure}

\subsection{Modulational instability of two coupled NLS equations}
\label{sec:MI_CNLS}
In this section we discuss the consequences of the results obtained so far in the framework of the study of the instabilities of a system of two coupled NLS equations.

In the scalar case, the focusing NLS equation
\begin{equation}
\label{NLS}
 u_{t}=i (u_{xx} -2 s|u|^2 u )\,,\quad s=-1
\end{equation}
has played an important role in modelling modulational instability of continuous waves \cite{BF1967,HO1972}. This unstable behaviour is predicted for this equation by simple arguments and calculations. It is therefore rather remarkable that, on the contrary, a nonlinear coupling of two NLS equations  makes the unstable dynamics of two interacting continuous waves  fairly richer than that of a single wave. Since the method of investigating the stability of two coupled NLS equations that we have presented in the previous sections requires integrability, we deal here with the integrable system (\ref{expVNLS}), also known as \emph{generalized Manakov model}, that we recall here for convenience:
\begin{align*}
 u_{1t}&=i [u_{1xx} -2 (s_1|u_1|^2 +s_2|u_2|^2)u_1 ]\\
 u_{2t}&=i [u_{2xx} -2 (s_1|u_1|^2 +s_2|u_2|^2)u_2 ] \,.
\end{align*}
This system is obtained by setting  $\mathbf{v}=S\mathbf{u}$ in (\ref{2vecteq}), where
\begin{equation}\label{matrixS}
S=\left (\begin{array}{cc} s_1 & 0  \\ 0 & s_2  \end{array} \right )\,,\quad S^2= \mathds{1}\,.
\end{equation}
The system of two coupled NLS equations is of interest in various physical contexts and  the investigation of the stability of its solutions deserves special attention.

Once a solution $ u_1(x,t)$, $u_2(x,t)$ has been fixed, the linearized equations (\ref{linQeq}) around this solution are
\begin{equation}\label{linVNLS}
\begin{array}{l}
\delta u_{1t}=i \{ \delta u_{1xx} -2[ (2s_1|u_1|^2 +s_2|u_2|^2)\delta u_1 +s_1u_1^2 \delta u_1^* +s_2 u_1 u_2^* \delta u_2 +s_2 u_1 u_2 \delta u_2^* ]\}\,\\
\delta u_{2t}=i \{ \delta u_{2xx} -2[ (s_1|u_1|^2+2s_2|u_2|^2)\delta u_2 +s_2u_2^2 \delta u_2^* +s_1 u_2 u_1^* \delta u_1 +s_1 u_2 u_1 \delta u_1^* ]\}\,.
 \end{array}
 \end{equation}
The two coupling constants $s_1$, $s_2$, if non-vanishing, are just signs, $s_1^2=s_2^2=1$, with no loss of generality. Thus the CNLS system (\ref{expVNLS}) models three different processes, according to the defocusing or focusing self- and cross-interactions that each wave experiences. These different cases are referred to as (D/D) if $s_1=s_2=1$, as (F/F) if $s_1=s_2=-1$ and as (D/F) in the mixed case $s_1s_2=-1$.

\begin{figure}[h!]
\begin{center}
\includegraphics[width=0.6\textwidth]{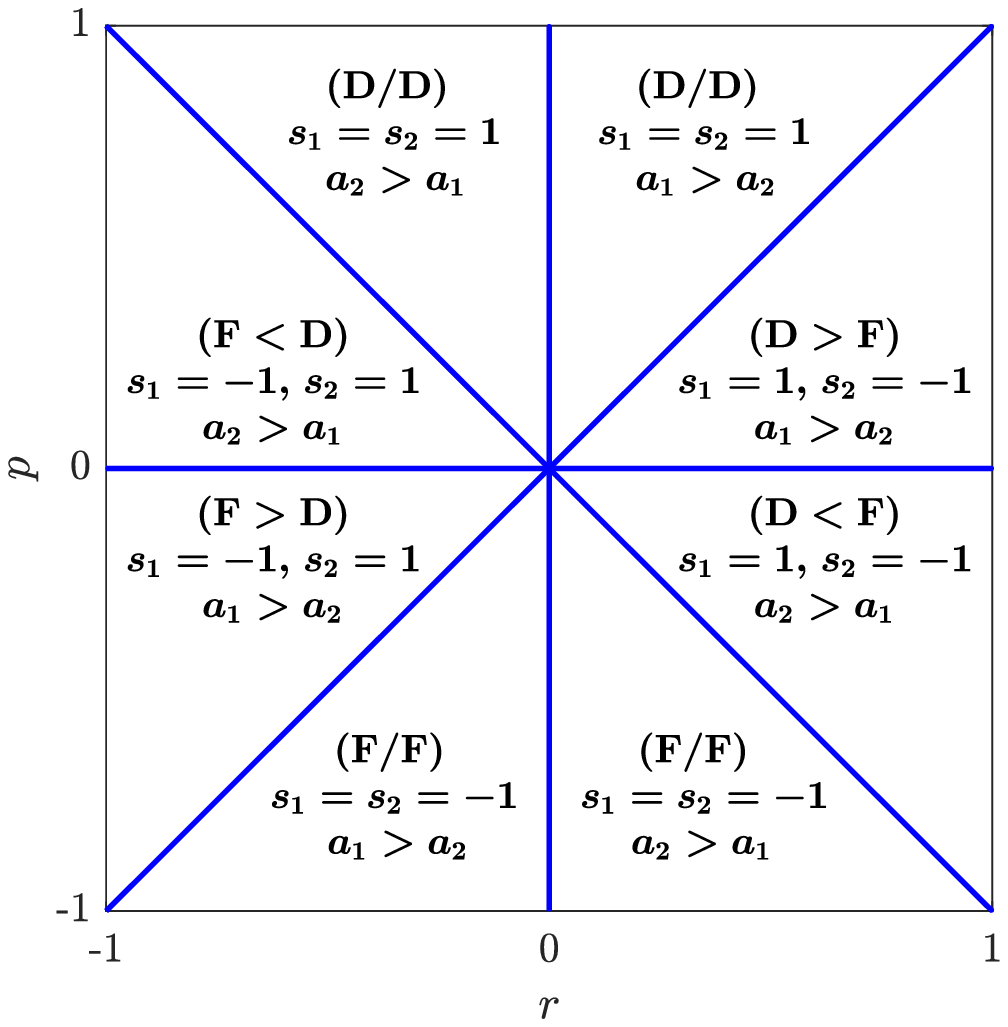}
\caption{The $(r,\,p)$-plane divided according to amplitudes $a_j$ and coupling constants $s_j$, $j=1,2$.}
\label{fig:foc_def_plane}
\end{center}
\end{figure}

Hereafter, our focus is on the stability of the CW solution (see (\ref{2vectsolut}), (\ref{vectora-b}) with $b_j=s_j a_j$)
\begin{equation}\label{2nlssolut}
 u_1(x,t)=a_1 e^{i(qx-\nu t)}\,,\quad
 u_2(x,t)= a_2 e^{-i(qx+\nu t)}\,,\quad
 \nu=q^2+2p \,,
\end{equation}
where the parameter $q$ is the relative wave number, which may be taken to be non negative $q\geq 0$, and $a_1$, $a_2$ are the two amplitudes, whose values, with no loss of generality, may be real and non negative, $a_j\geq 0$.
As for the notation, the parameter $p$ is defined by (\ref{parap}) which, in the present reduction,  reads
\begin{subequations}\label{VNLSparam}
\begin{equation}
\label{VNLSpparam}
p= s_1 a_1^2+s_2 a_2^2  \,.
\end{equation}
To make contact with the stability analysis presented in the two preceding sections, we introduce also the second relevant parameter $r$ (\ref{parar}) whose expression, in the present context, is
\begin{equation}
\label{VNLSrparam}
r= s_1 a_1^2-s_2 a_2^2 \,.
\end{equation}
\end{subequations}
At this point, we show in Figure \ref{fig:foc_def_plane} the $(r,\,p)$-plane as divided into octants, according to different values of amplitudes and coupling constants.

We note that, as pointed out by Proposition \ref{prop:6}, it is sufficient to confine our discussion to the half-plane $r\geq 0$, where four different experimental settings are represented, one in each of these four octants. These four states of two periodic waves can be identified either by the parameters $p$ and $r$ or by amplitudes and coupling constants according to the relations
\begin{equation}
\label{paramrelat}
s_1 a_1^2 =\frac12(p+r)\,,\quad s_2 a_2^2 =\frac12(p-r) \,,
\end{equation}
each octant being characterized as shown in Figure \ref{fig:foc_def_plane}.
In particular the octant $r>p>0$ will be referred to as $(\mathrm{D}>\mathrm{F})$, whereas the octant $-r<p<0$ as $(\mathrm{D}<\mathrm{F})$, to indicate  the wave with larger amplitude. Moreover, we observe again that setting $q=0$, which characterizes two standing waves with equal wavelength, makes the stability properties of the CW solution very similar to that of the NLS equation. Indeed, in this  case solution (\ref{2nlssolut}) is stable if both waves propagate in a defocusing medium, $s_1=s_2=1$, and is unstable in the opposite case $s_1=s_2=-1$. However, this solution remains stable even if the wave $u_2$ feels a self-focusing effect, \textit{i.e.} $s_2=-1$, provided the other wave $u_1$ in the defocusing medium, $s_1=1$, has larger amplitude $a_1>a_2$. Similarly, when $q=0$, the CW solution is unstable if $s_1=1$ and $s_2=-1$, and if the larger amplitude wave is the one which propagates in the focusing medium, $a_2>a_1$. These remarks follow from explicit formulae (\ref{q0kz}) and (\ref{VNLSpparam}), and they are clearly evident in Figure \ref{fig:foc_def_plane} where, for $r>0$, the upper two octants correspond to $p>0$, and the lower two to $p<0$.

The limit $q\rightarrow 0$ of the results presented in the previous sections should be considered as formally singular. Indeed, the behaviour of the CW (\ref{2nlssolut}) against small generic perturbations becomes significantly different from that reported above if the wave number mismatch $2q$ is non vanishing. For one aspect, this is apparent from our first
\begin{Remark}
\label{rem:1} If $q\neq 0$, then for no value of the amplitudes $a_1$, $a_2$ with $a_1a_2\neq 0$, and for no value of the coupling constants $s_1$, $s_2$, the spectrum $\textbf{S}_x$ is entirely real.
\end{Remark}
This follows from Proposition \ref{prop:9} and from our classification of the spectra in the parameter space into five types, all of which have a non-empty complex (non-real) component. The condition $a_1 a_2 \neq 0$ for this statement to be true coincides with the condition that the point $(r,p)$ neither belongs to the threshold curve $p=p_+(r)$, nor to the threshold curve $p=p_-(r)$, see (\ref{VNLSparam}). In particular, in the case $p=r>0$, which is equivalent to setting $a_2=0$ and $s_1=1$, the roots of $P_W(w;\lambda)$, (\ref{polyW}), can be given in closed form,
\[
w_1= -(1/2)q +\sqrt{(\lambda+q/2)^2-a_1^2}\,,\quad w_2= -(1/2)q -\sqrt{(\lambda+q/2)^2-a_1^2}\,,\quad w_3= -\lambda+q\,,
\]
so that
\begin{align*}
& k_1=w_2-w_3=\lambda-\frac32 q -\sqrt{(\lambda+q/2)^2-a_1^2}\,,\nonumber\\
& k_2=w_3-w_1=-\lambda + \frac32 q -\sqrt{(\lambda+q/2)^2-a_1^2}\,,\nonumber\\
& k_3=w_1-w_2=2\sqrt{(\lambda+q/2)^2-a_1^2} \,,
\end{align*}
with the implication that $\lambda$ has to be real to be in $\textbf{S}_x$. In this case the spectrum is not immediately identifiable as one of the five types of spectra described in Proposition \ref{prop:9}, for it corresponds to a threshold case between two of such types, and features one gap, no branches, and no loops. Moreover, this spectrum coincides with that obtained for the defocusing NLS equation via a Galilei transformation. In a similar fashion, if the point $(r,p)$ is taken on the curve $p=p_-(r)$, namely $p=-r<0$ or, equivalently, $a_1=0, s_2=-1$, the corresponding spectrum is that of the CW solution of the focusing NLS equation modulo a Galilei transformation, namely with no gap on the real axis, one branch on the imaginary axis, and no loop. Remark \ref{rem:1} has a straight implication on the stability of solution (\ref{2nlssolut}). This can be formulated as
\begin{Remark}
\label{rem:2}
If  $q\,a_1a_2 \neq 0$ then the CW solution (\ref{2nlssolut}) is unstable.
\end{Remark}
The relevant point here is the dependence on $\lambda $ of the frequencies
$\omega_j(\lambda)$ over the spectrum $\textbf{S}_x$. As we have shown in the previous section, if $q\,a_1a_2 \neq 0$, the spectrum $\textbf{S}_x$ consists of two components: one, $\textbf{RS}_x$, is the real axis $\mathrm{Im}({\lambda})=0$, with possibly one or two gaps  (see Proposition \ref{prop:8}), and the second one, $\textbf{CS}_x$,  consists of branches and/or loops where $\lambda$ runs off the real axis (see Proposition \ref{prop:9}). Therefore the spectral representation (\ref{lincomb_Sx}) of the perturbation
\begin{equation}\label{deltau}
\mathbf{\delta u}=\left (\begin{array}{c} \delta u_1 \\ \delta u_2  \end{array} \right )
\end{equation}
takes the form
\begin{equation}\label{spectrdeltau}
\begin{array}{lcl}
\mathbf{\delta u} &=& e^{i(qx\sigma_3-\nu t)} \left \{ {\displaystyle \int_{\textbf{RS}_x} d\lambda} \sum_{j=1}^3 \left [ e^{i(xk_j- t\omega_j)} \mathbf{f}^{(j)}_+(\lambda)+e^{-i(xk_j- t\omega_j)} \mathbf{f}^{(j)}_- (\lambda)\right ] +\right.\\
&& \hspace{2.62cm}+  \left. {\displaystyle \int_{\textbf{CS}_x} d\lambda \left [ e^{i(xk_3- t\omega_3)} \mathbf{f}^{(3)}_+(\lambda)+e^{-i(xk_3- t\omega_3)} \mathbf{f}^{(3)}_- (\lambda)\right ] } \right \}\,,
\end{array}
\end{equation}
which is meant to separate the contribution to $\mathbf{\delta u}$ due to the real part of the spectrum from that coming instead from the complex values of the integration variable $\lambda$, this being the integration over branches and loops. The 2-dim vector functions $\mathbf{f}^{(j)}_{\pm}(\lambda) $ do not play a role here and are not specified. On the contrary, the reality property of the frequencies $\omega_j$ over the spectrum is obviously essential to stability. Since the reality of the frequencies $\omega_1$, $\omega_2$, $\omega_3$  for $\lambda$ real has been proved in Proposition \ref{prop:5}, it remains to show that indeed $\omega_3(\lambda)$ is not real if $\lambda$ belongs to branches or loops. This follows from the explicit expression (\ref{eigenFREQ}), namely
\begin{equation}
\label{GAIN}
\omega_3= k_3(w_3- \lambda) = \Omega+i\Gamma\,,
\end{equation}
where $k_3$ is real but $w_3$ and $\lambda$ cannot be  real, and cannot have same imaginary part as proved in Proposition \ref{prop:5}, (see Section \ref{subsec:branchesandloops}).
Here the imaginary part $\Gamma$ of $\omega_3$ defines the \emph{gain function} over the spectrum. This (possibly multivalued) function of $k_3$ plays an important role in the initial stage of the unstable dynamics. Precisely, its dependence on the wave number $k_3$ gives important information on the instability band and on timescales \cite{DLS2017}.

The physical interpretation is more transparent if these considerations are stated in terms of the amplitudes  of the CW solution (\ref{2nlssolut}) for the three choices of the coupling constants $s_1$, $s_2$ corresponding to integrable cases. In this respect, it is convenient to translate the classification of the spectra in the $(a_1\,,\,a_2)$-plane, in particular, and with no loss of generality, in the quadrant $a_1\geq 0$, $a_2\geq 0$.
The four threshold curves $p_{\pm}(r)$, $p_S(r)$, $p_C(r)$ which have been found in the $(r,p)$-plane, see Figure  \ref{fig:spectrum}, are reproduced below in the $(a_1,a_2)$-plane, see Figure \ref{fig:a_1,a_2-plane}, according to the coordinate transformations (\ref{VNLSparam}) and (\ref{paramrelat}). The outcome of this analysis is summarized by the following

\begin{figure}[h!]
\centering
\begin{subfigure}[b]{0.33\textwidth}
\includegraphics[height=4.5cm]{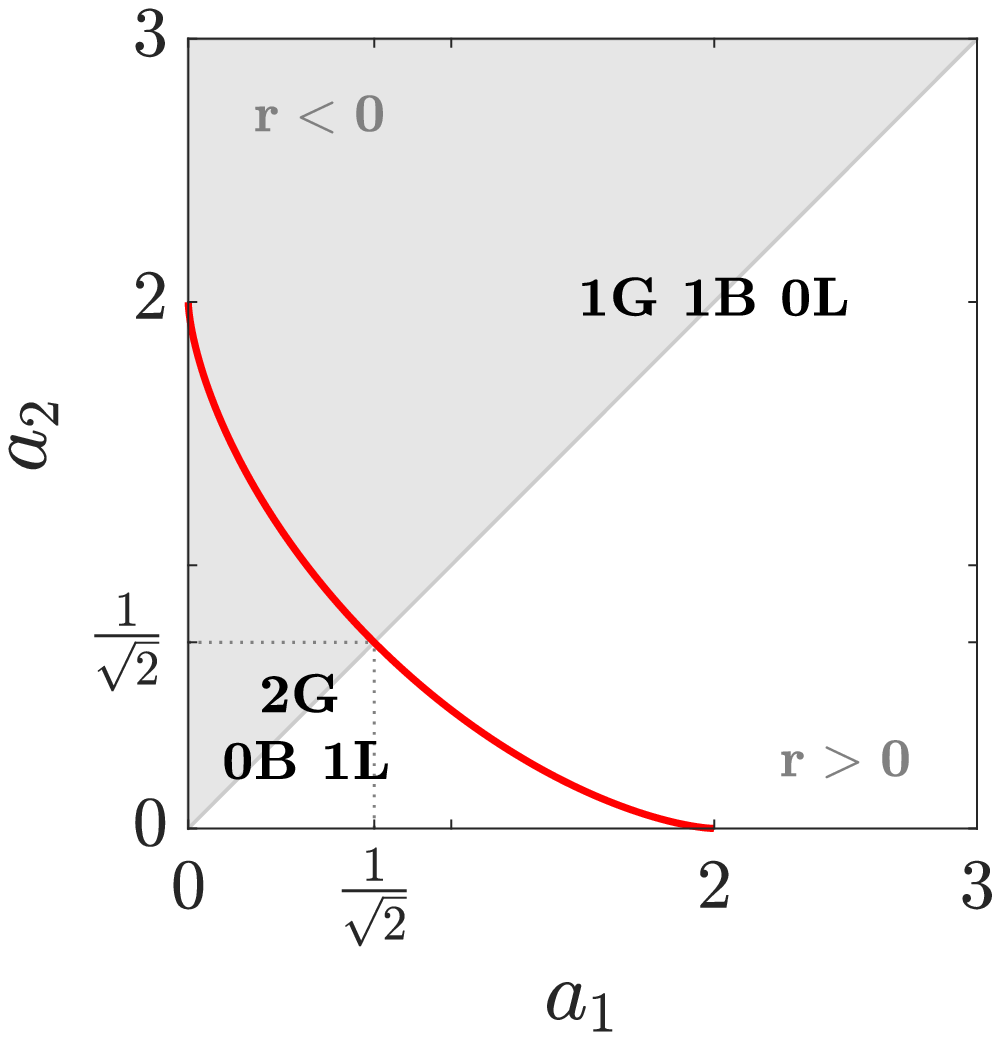}
\caption{(D/D) $s_1=s_2=1$\label{fig:s1=1_s2=1}}
\end{subfigure}
\hspace{-0.4cm}
\begin{subfigure}[b]{0.33\textwidth}
\includegraphics[height=4.5cm]{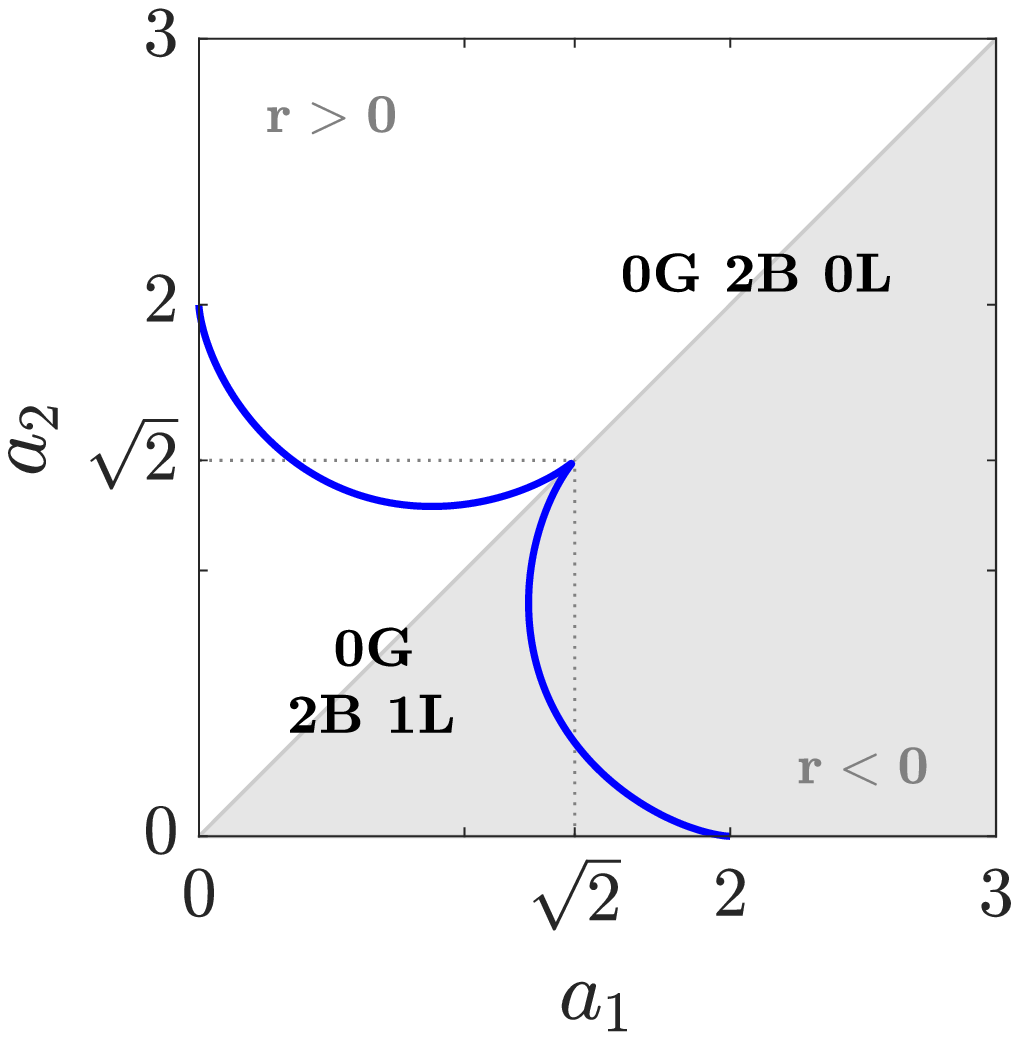}
\caption{(F/F) $s_1=s_2=-1$ \label{fig:s1=-1_s2=-1}}
\end{subfigure}
\hspace{-0.2cm}
\begin{subfigure}[b]{0.33\textwidth}
\includegraphics[height=4.5cm]{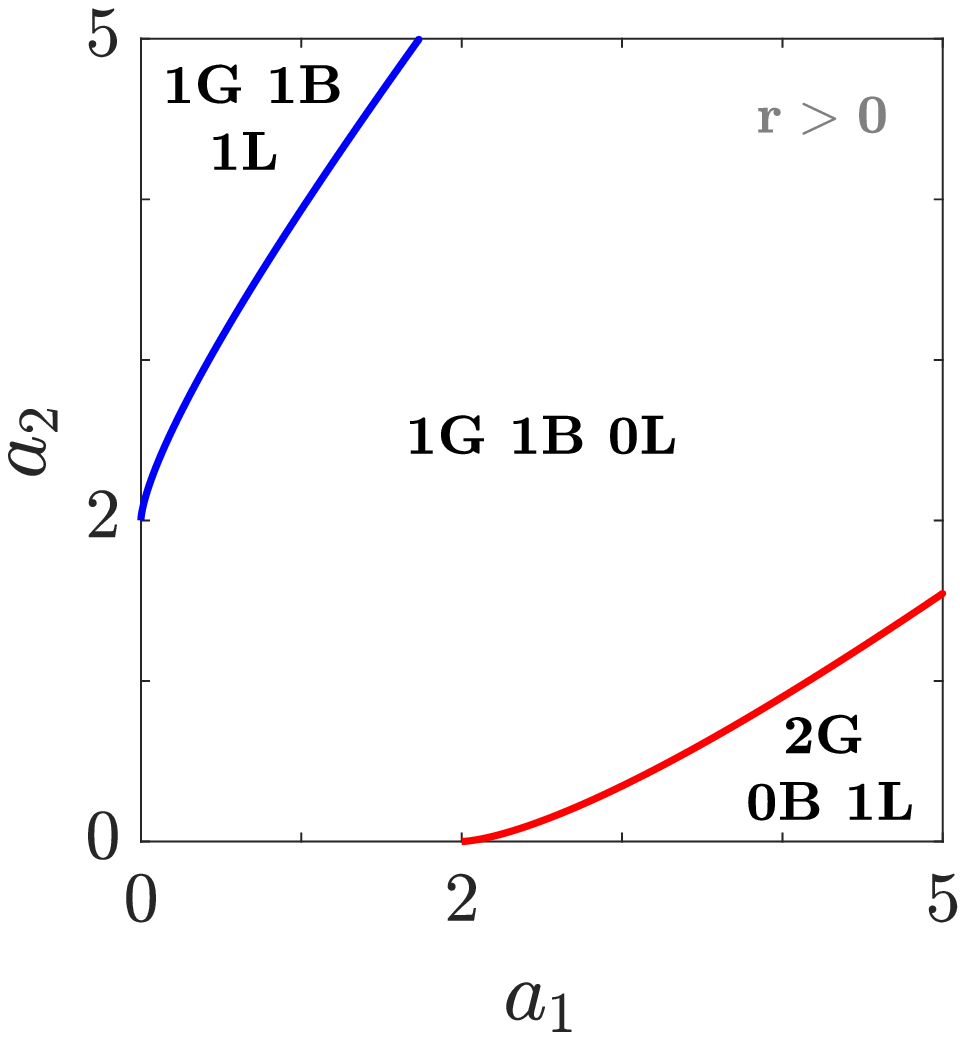}
\caption{(D/F) $s_1=1\,,\,s_2=-1$\label{fig:s1=1_s2=-1}}
\end{subfigure}
\caption{The $(a_1,a_2)$-plane (see Proposition \ref{prop:10}).\label{fig:a_1,a_2-plane}. In Figures \ref{fig:s1=1_s2=1} and \ref{fig:s1=-1_s2=-1}, gray portions correspond to $r<0$.}
\end{figure}

\begin{Proposition}
\label{prop:10} For each of the three integrable cases of the two coupled NLS equations (\ref{expVNLS}) the spectrum is of the following  types.
 \begin{description}
 \item $\mathrm{(D/D)}$ $s_1=s_2=1$ (see Figure \ref{fig:s1=1_s2=1})
 \newline
The lower octant $(a_1\geq a_2)$, which corresponds to $r\geq 0$, gets divided in two parts by the threshold curve $p_S$ which intersects the line $a_2=a_1$ at the point $(1/\sqrt{2}, 1/\sqrt{2})$. In the lower, finite, part of this octant the spectrum $\textbf{S}_x $ is of type $\mathrm{2G\,\, 0B\,\, 1L}$, while in the other part it is of type $\mathrm{1G\,\, 1B\,\, 0L}$.
 \item $\mathrm{(F/F)}$ $s_1=s_2=-1$ (see Figure \ref{fig:s1=-1_s2=-1})
 \newline
 The upper octant $(a_2\geq a_1)$, which corresponds to $r\geq 0$, gets divided in two parts by the threshold curve $p_C$ which intersects the line $a_2=a_1$ at the point $(\sqrt{2}, \sqrt{2})$. In the lower, finite, part of this octant the spectrum $\textbf{S}_x $ is of type $\mathrm{0G\,\, 2B\,\, 1L}$, while in the other part it is of type $\mathrm{0G\,\, 2B\,\, 0L}$.
 \item $\mathrm{(D/F)}$ $s_1=1\,,\,s_2=-1$ (see Figure \ref{fig:s1=1_s2=-1})
 \newline
The whole quadrant corresponds to $r\geq 0$. It is divided into three infinite portions by the two threshold curves $p_C$, in the upper octant, and $p_S$ in the lower one. The spectrum is of type $\mathrm{1G\,\, 1B\,\, 1L}$ in the upper part, it is of type $\mathrm{1G\,\, 1B\,\, 0L}$ in the middle part, and of type $\mathrm{2G\,\, 0B\,\, 1L}$ in the lower part.
\end{description}
\end{Proposition}
These statements are straight consequences of Proposition \ref{prop:9} via  transformation (\ref{paramrelat}).

We conclude by mentioning the limit case $r=0$, see the end of the previous section. This concerns only the (D/D) and (F/F) cases of Proposition \ref{prop:9} and it coincides with the limit $a_1=a_2$, namely with the case in which the two wave amplitudes are strictly equal.
On this particular line of the $(a_1, a_2)$-plane, \textit{i.e.} $r=0$ in the $(r, p)$-plane, the spectra are of four types, according to numbers of gaps, branches and loops, as it has been shortly discussed in Subsection \ref{subsec:branchesandloops}.

\section{Summary and conclusions}
\label{sec:conclusion}
A sufficiently small perturbation of a solution of a (possibly multi-component) wave equation satisfies a linear equation. If the wave equation is integrable, the solution of this linear equation is formulated in terms of a set of eigenmodes whose expression is explicitly related to the solutions of the Lax pair.  We give this connection in a general $N \times N $ matrix formalism, which is local (in $x$), and does not require specifying the boundary condition nor the machinery of the direct and inverse spectral problem. A by-product of our approach is the definition of the spectrum $\mathbf{S}_x$, associated with the unperturbed solution of the wave equation. It is worth stressing that the spectrum $\textbf{S}_x$ for $N>2$ does not coincide with the spectrum of the Lax equation $\Psi_x=(i\lambda \Sigma+Q)\Psi$ in the complex $\lambda$-plane. This result is explicitly shown for $N=3$ in the instance of continuous wave solutions of two CNLS equations. In this case, we define as functions of $\lambda$  on the spectrum $\textbf{S}_x$ the set of wave numbers $k_j(\lambda)$ and frequencies $\omega_j(\lambda)$ with the implication that the dispersion relation is given in parametric form, the parameter being the spectral variable $\lambda$ which appears in the Lax pair. In general, the spectrum is a complicated piecewise  continuous curve.  It obviously changes in the parameter space which is the set of values of the amplitudes of the two CWs, the mismatch $q$ of their wave numbers, and the values of the coupling constants $s_1$, $s_2$.  Apart from particular values of the parameters, the computation of the spectrum is not achievable in analytic form, and has to be done numerically. The knowledge of the spectrum is sufficient to assess the stability of the CW solution. Generally, the spectrum consists of the real $\lambda$-axis with possibly one or two forbidden bands (gaps) and few additional finite curves which may be open (branches) or closed (loops). According to these topological properties, spectra can be classified in five different types to completely cover the entire parameter space. Only few marginal cases require separate consideration. Physically relevant information comes from the $\lambda$ dependence of the eigenfrequency on branches and loops. In particular, one can read out of this dependence on $\lambda$  the instability band, whether at large (as in the Benjamin-Feir instability) or at small wavelengths, and the time scale of the exponential growth in time of the perturbation. This is  characterized by the imaginary part of the complex frequency, namely by the gain function. However, the investigation of this interesting aspect of the stability analysis is not reported here as it requires further analysis and computations. This part of our work will be reported elsewhere \cite{DLS2017}.

\appendix
\numberwithin{equation}{section}
\section{Structure of the gaps of the spectrum $\mathbf{S}_x$}
\label{app:A}
The aim of this section is to provide additional details of the gap structure of the spectrum $\mathbf{S}_x$ and to give a proof of Proposition \ref{prop:8} in Section \ref{subsec:gaps}, which states that the real part of the  spectrum $\textbf{S}_x$ has either one, two or no gaps in the $(r, p)$-plane with $r\geq 0$.
The no gap case occurs if the three eigenvalues $w_j(\lambda)$ are real on the entire $\lambda$-axis. This happens if the discriminant (\ref{discrW}) is positive for any real value of $\lambda$. By varying the value of $p$ and/or $r$ a gap may open up at the double zeros of the discriminant $D_W(\lambda;q,p(\lambda),r)$. Inside a gap the discriminant is negative and the three wave-numbers $k_j(\lambda)$ (\ref{k}) have a non vanishing imaginary part.
We define implicitly the function $p\equiv p(\lambda)$ as those values of the parameter $p$ such that the discriminant $D_W(\lambda;q,p(\lambda),r)=0$.

\begin{figure}[h!]
\centering
\begin{subfigure}[h]{0.48\textwidth}
\includegraphics[width=\textwidth]{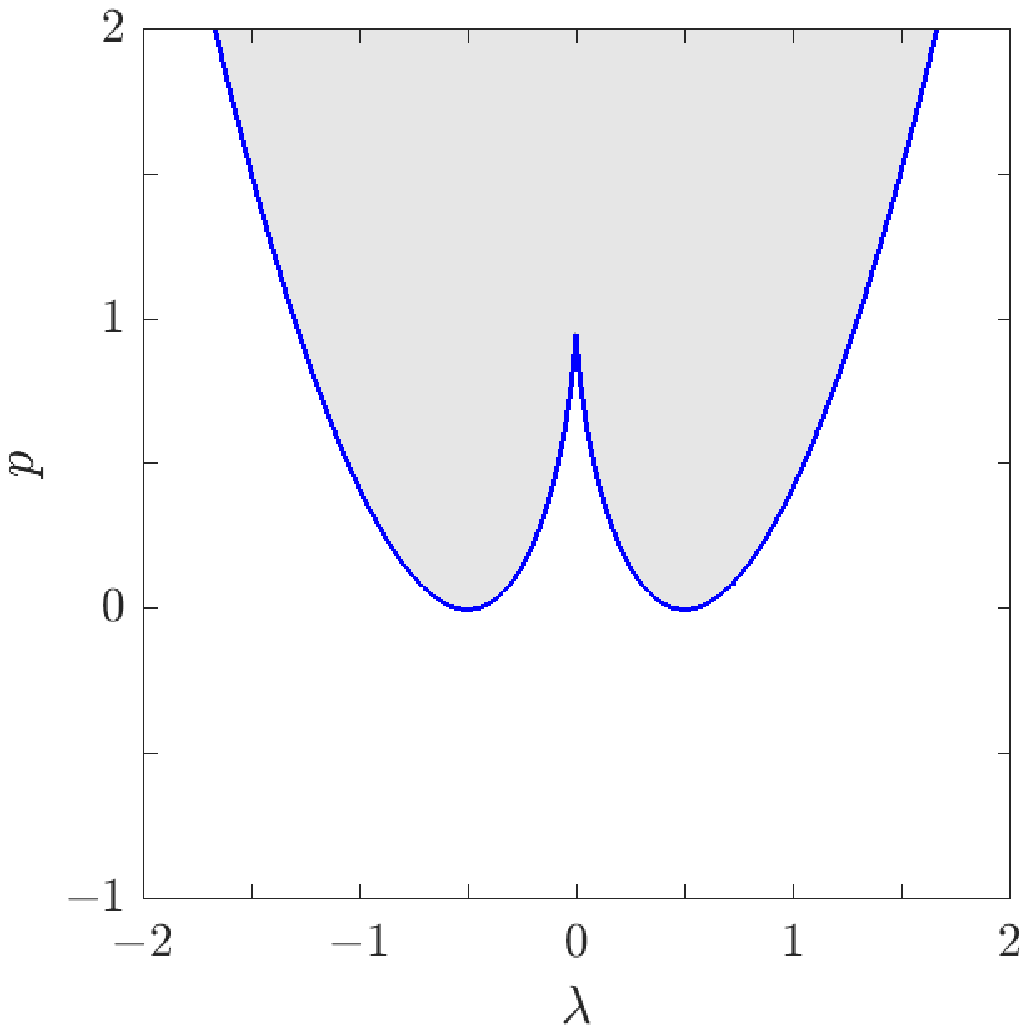}
\caption{$r=0$, $q=1$\label{fig:gapsa}}
\end{subfigure}
\;
\begin{subfigure}[h]{0.48\textwidth}
\includegraphics[width=\textwidth]{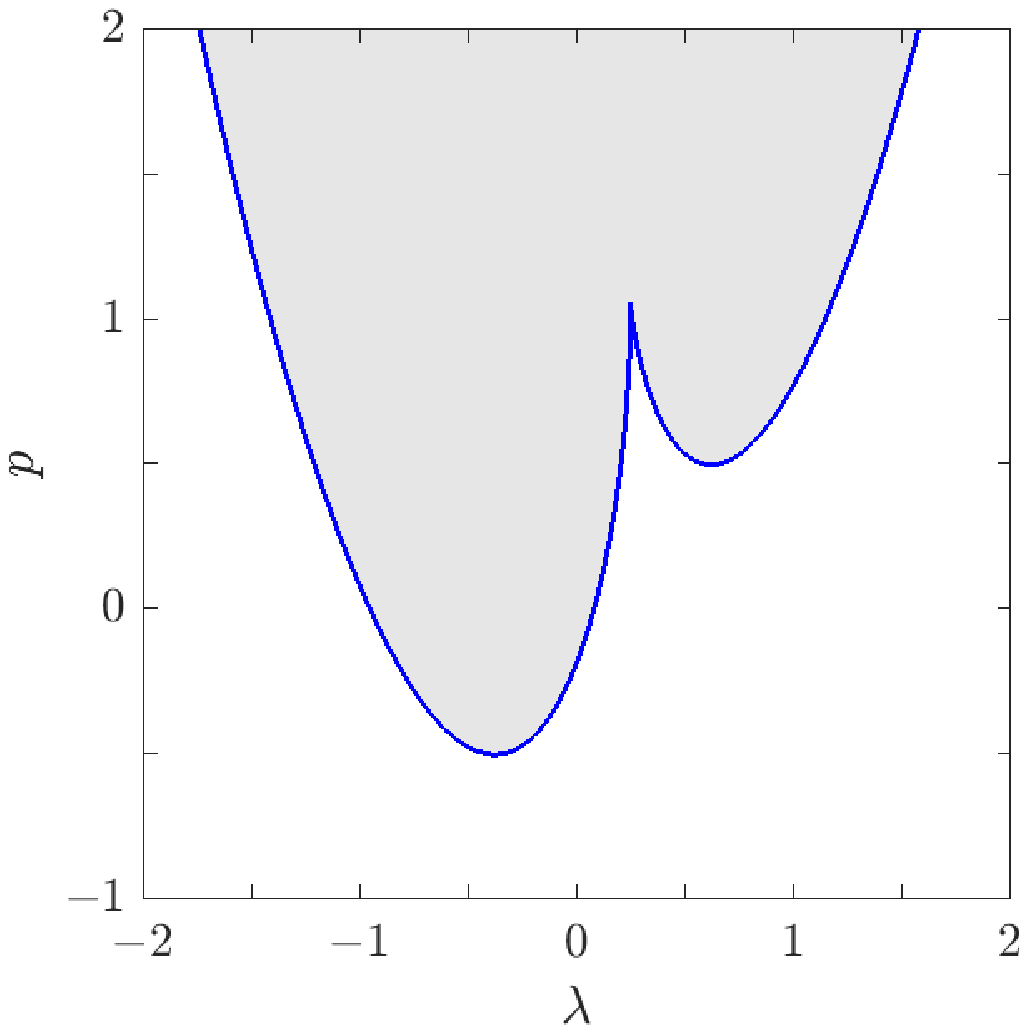}
\caption{$r=0.5$, $q=1$\label{fig:gapsb}}
\end{subfigure}
\begin{subfigure}[h]{0.48\textwidth}
\includegraphics[width=\textwidth]{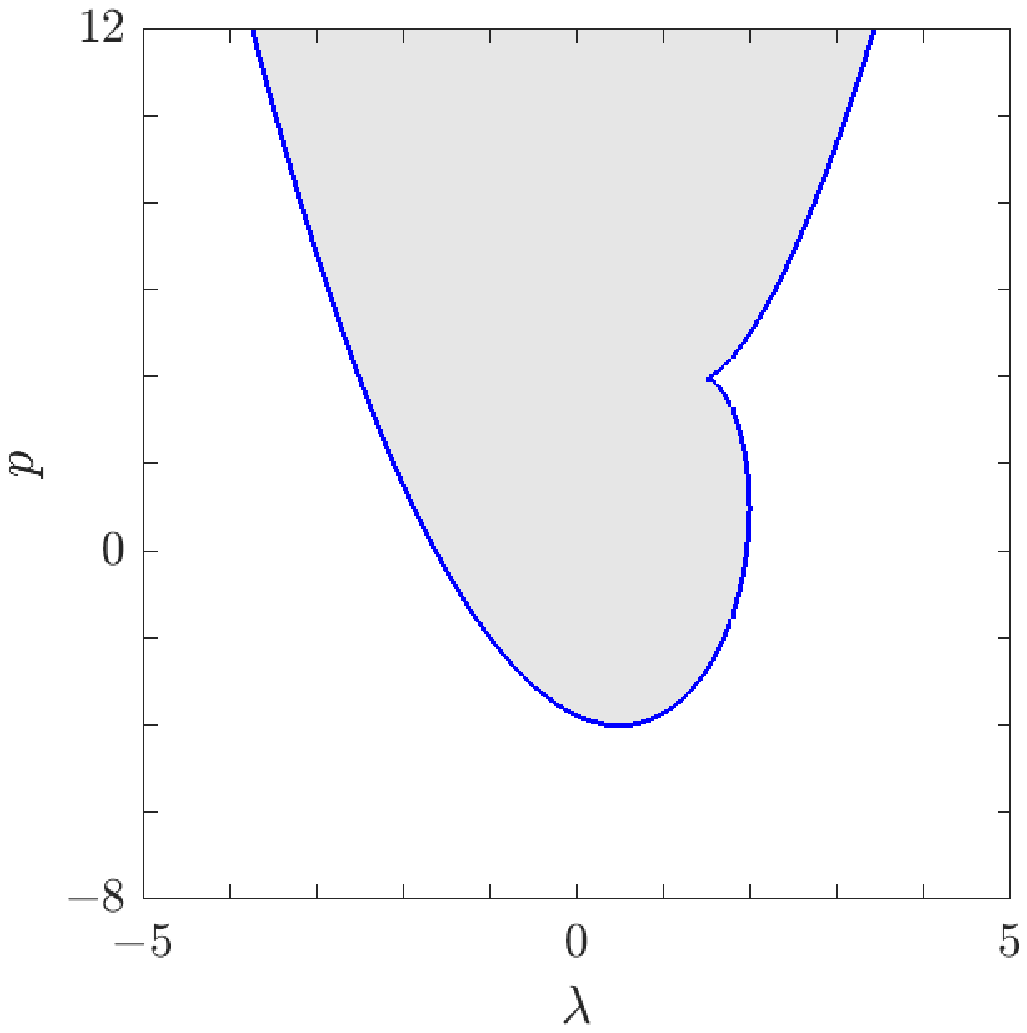}
\caption{$r=4$, $q=1$\label{fig:gapsc}}
\end{subfigure}
\;
\begin{subfigure}[h]{0.48\textwidth}
\includegraphics[width=\textwidth]{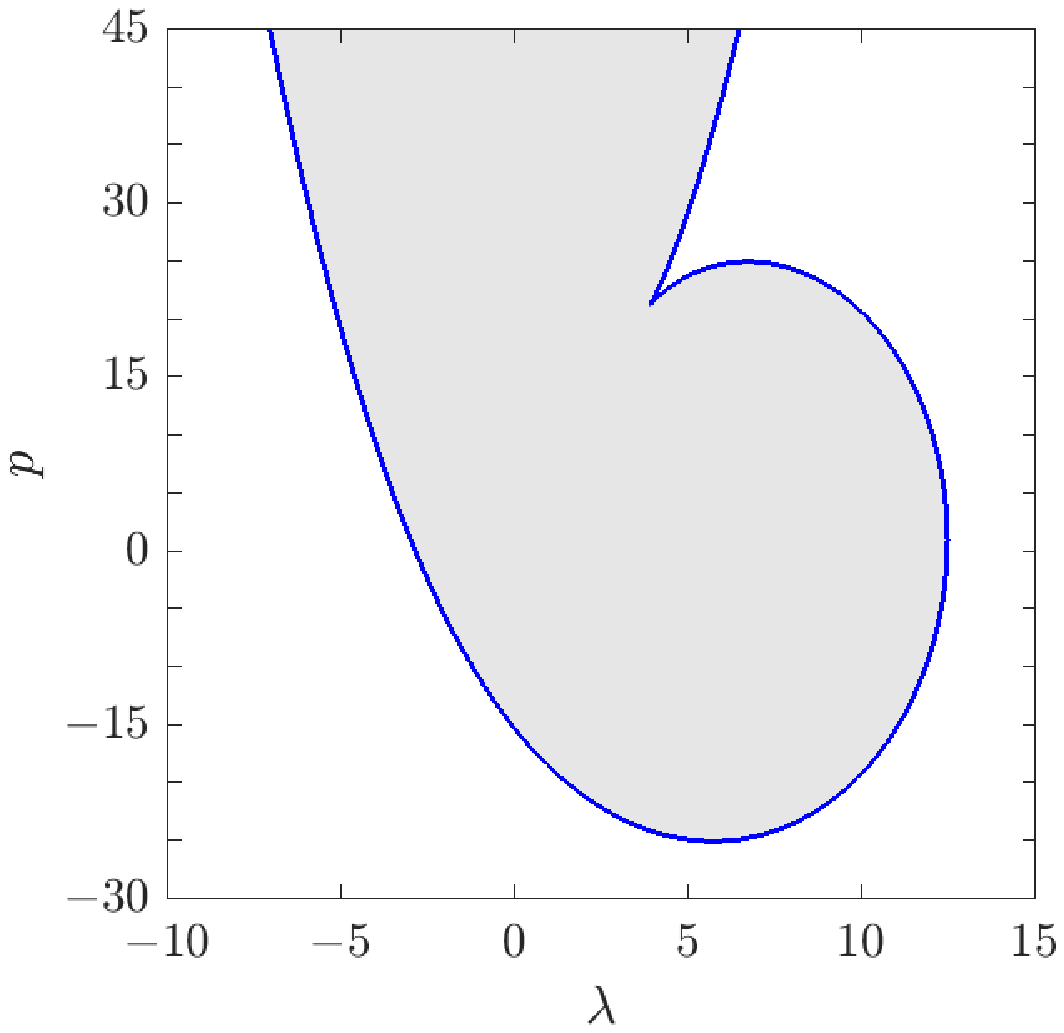}
\caption{$r=25$, $q=1$\label{fig:gapsd}}
\end{subfigure}
\caption{$p(\lambda)$ defined by $D_W(\lambda;q,p(\lambda),r)=0$; the grey area corresponds to $D_W<0$.\label{fig:gaps}}
\end{figure}

In order to compute the number of gaps from expression (\ref{discrW}) of the discriminant $D_W(\lambda;q,p,r)$, we plot $p(\lambda)$ as a function of $\lambda$ real, for few fixed values of $r$, and $q=1$.
Four such plots of $p(\lambda)$ are shown in Figure \ref{fig:gaps} for four different values of $r$. The intersections of this plotted curve with the straight line corresponding to a constant value of $p$ provide the endpoints of those gaps which occur at the given value of $p$ (and at the fixed values of $r$ and $q$). For instance, the analytical results presented above for $r=0$, see Proposition \ref{prop:7}, can be clearly read out of the left plot in Figure \ref{fig:gapsa}. The two local minima at $\lambda =\pm 0.5$,  $p=0$ show the opening of the two gaps that we have analytically found for $0<p<1$. For future reference, we also note that the local maximum of $p(\lambda)$ is taken at $\lambda=\lambda_S=0$, namely $p(\lambda_S)=p_S=1$, and that the function $p(\lambda)$ has a cusp (a double singular point), namely is not analytic, at $\lambda_S$. If the parameter $r$ increases, $r>0$, this plot gets asymmetrically deformed as shown in Figures \ref{fig:gapsb}, \ref{fig:gapsc} and \ref{fig:gapsd}.
Figure \ref{fig:gapsb} shows that no gap exists for $p< -r=-0.5$, while for $ -0.5=-r<p<r=0.5$ only one gap occurs. Instead, for $0.5=r<p<p_S$ two distinct gaps appear. Finally, for $p_S < p < +\infty$ again only one gap is present. As in the previous plot, here $p_S$ denotes the local maximum and singular point of $p(\lambda)$. Moreover, we observe that this graph is consistent with the following alternative expression of the discriminant (\ref{discrW}) for $p=-r$,
\[
 D_W(\lambda,q,-r,r)= (4q\lambda+2q^2-r)^2 [(2\lambda-q)^2+4r]\,,
\]
which is clearly positive definite for any non negative $r$, and any real $\lambda$ with the exception of
$\lambda= \frac{r}{4q}-\frac{q}{2}$ where $p(\lambda)$ takes its minimum value.

By increasing further the value of $r$, Figure \ref{fig:gapsb} changes and eventually looks different as the local maximum $p_S$ becomes a local minimum. The transition is illustrated in Figure \ref{fig:gapsc} computed at $r=4$, with $q=1$. After the transition, the curve $p(\lambda)$ takes the form shown in Figure \ref{fig:gapsd} computed at $r=25$, with $q=1$. Again, $p=-r$ is still the minimum of $p(\lambda)$ while $p=r$ is instead a local maximum.
The singularity at $\lambda=\lambda_S$ is such that now $p_S=p(\lambda_S) $ is a local minimum with $p_S<r$.
Indeed, $\lambda_S$ is the only singularity of $p(\lambda)$ and its dependence on the parameter $r$, $\lambda_S=\lambda_S(r)$,  is implicitly expressed by its inverse relation $r=r(\lambda_S)$,
\begin{equation}
\label{singular}
 r=2\frac{\lambda_S}{q}\left(q^2+\frac{4}{27}\lambda_S^2\right)\,.
 \end{equation}
This yields a one-to-one correspondence between any non negative real $r$ and a real $\lambda_S(r)$ since we prove that, for any $r\geq 0$, the cubic equation (\ref{singular}) has two complex conjugate roots. Moreover, we find also the following useful relations associated with the singular value $\lambda_S$:
\begin{equation}
\label{singrel_app}
 p_S=q^2+\frac{4}{3}\lambda_S^2\,;\quad
 w_j=-\frac13 \lambda_S\,,\quad
 z_j=-q^2+\frac29 \lambda_S^2\,,\quad
 k_j=\omega_j=0\,,\;j=1,2,3\,.
 \end{equation}
 The transition between the two regimes with $p_S(r)>r$ and $p_S(r)<r$ takes place at the special threshold value $r=r_T$, where $p_S(r_T)=r_T$. With self-evident notation, we find that
\begin{equation}
\label{trans}
 r_T= p_T=p_S(r_T)=4q^2 \,,\quad \lambda_T=\frac{3}{2}q\,.
\end{equation}
The implicit expressions of $\lambda_S(r)$ (\ref{singular})
is made explicit in (\ref{expli}). The function $p_S(r)$ is plotted in Figure \ref{fig:pS(r)}.

\section{Analysis of the $x$-spectrum}
\label{app:B}
As pointed out in Section \ref{subsec:branchesandloops}, the nature and the number of the components of the $x$-spectrum $\mathbf{S}_{x}$, seen as a curve in the $\lambda$-plane, are classified by the number of sign-changes of the discriminant of $\mathcal{P}(\zeta;\lambda,q,p,r)$ with respect to $\lambda$, \textit{i.e.} $\Delta_{\lambda}\mathcal{P}(\zeta;\lambda,q,p,r)=\mathcal{Q}(\zeta;q,p,r)$, for $\zeta\geq 0$ (see also (\ref{eq:QQ}) in the following).
The aim of this section is to give details of this result. Without loss of generality, from now on we will assume $q=1$.

Given the (cubic) polynomial $P_{W}(w;\lambda,q,p,r)$, we would like to describe the  geometrical locus in the $\lambda$-plane constituted by the complex values of $\lambda$ such that, for each value of $\lambda$ in the locus, the polynomial $P_{W}(w;\lambda,q,p,r)$ has at least two $w$-roots such that their difference is real.
As discussed in Section \ref{subsec:branchesandloops},
since we are interested in the differences of the roots, it is convenient to introduce the polynomial $\mathcal{P}(\zeta;\lambda,q,p,r)$, see (\ref{polyK}), whose $\zeta$-roots are the squares of the differences of the original $w$-roots of $P_{W}(w;\lambda,q,p,r)$.
The polynomial $\mathcal{P}(\zeta;\lambda,q,p,r)$ can be conveniently rewritten as
$$
\mathcal{P}(\zeta;\lambda,q,p,r)=
\left(\begin{array}{ccccc}1 & \lambda & \lambda^{2} & \lambda^{3} & \lambda^{4}\end{array}\right)\cdot
\Pi
\cdot\left(\begin{array}{c}1\\\zeta\\\zeta^{2}\\\zeta^{3}\end{array}\right)
$$
with
$$
\Pi\equiv \Pi(p,r)=
\left(\begin{array}{cccc}
    4\,(p-1)^{3}+27\,r^{2} & 9\,(p-1)^{2} & 6\,(p-1) & 1\\
    -36\,(p+2)\,r & 0 & 0 & 0\\
    4\,(8+20\,p-p^{2}) & -24\,(p-1) & -8 & 0\\
    32\,r & 0 & 0 & 0\\
    -64 & 16 & 0 & 0
\end{array}\right)\,.
$$
Observe that $\zeta=k_{j}^{2}$ for some $j$. As  in Section \ref{subsec:branchesandloops}, let $j=3$. The polynomial $\mathcal{P}(\zeta;\lambda,q,p,r)$ can be regarded as a $3^{\mathrm{rd}}$ degree polynomial in $\zeta$, as well as a $4^{\mathrm{th}}$ degree polynomial in $\lambda$. Hence,  our problem translates into finding
the complex values of $\lambda$ for which $\mathcal{P}(\zeta;\lambda,q,p,r)$, seen as a polynomial in $\zeta$, admits at least one real, positive root $\zeta=k_{3}^{2}$.
The $x$-spectrum $\mathbf{S}_{x}$ then coincides with the locus of the roots of $\mathcal{P}(\zeta;\lambda,q,p,r)$ regarded as a polynomial in $\lambda$ for all values of $\zeta > 0$, namely for $\zeta$ considered as a real, positive parameter.

From the algebraic-geometric point of view, the locus of the roots of $\mathcal{P}(\zeta;\lambda,q,p,r)$ in the $\lambda$-plane is an algebraic curve; this can be given implicitly as a system of two polynomial equations in two unknowns by setting $\lambda=\mu+i\,\rho$ and then separating the real and the imaginary parts of $\mathcal{P}(\zeta;\lambda,q,p,r)$.
In the following, we will apply Sturm's chains (\textit{e.g.}, see \cite{demidovich1981computational,HM1990}) on $\mathcal{Q}(\zeta;q,p,r)$, the discriminant of $\mathcal{P}(\zeta;\lambda,q,p,r)$, and invoke Sturm's theorem to study its roots, in order to obtain a classification of the different loci $\mathbf{S}_x$.

\subsection{Sturm chains  and spectra classification in the $(r,p)$-plane}
Let $P(x)$ be a polynomial in $x$ with real coefficients, and let $\text{deg}(P)$ be its degree. A Sturm's chain \cite{demidovich1981computational,HM1990}
\[
\begin{aligned}
& P^{(0)}(x)=P(x), \\
& P^{(1)}(x)=\frac{\mathrm{d}}{\mathrm{d}x}P^{(0)}(x) \\
& P^{{(2)}}(x)=-\text{Remainder}(P^{(0)}(x),P^{(1)}(x)),\\
&\vdots\\
& P^{(j)}(x)=-\text{Remainder}(P^{(j-2)}(x),P^{(j-1)}(x))\,,
\end{aligned}
\]
is a sequence of polynomials associated with a given polynomial $P(x)$, and its derivative, where the notation $\text{Remainder}(P_1,P_2)$ stands for the remainder of the polynomial division between $P_1$ and $P_2$.
By Sturm's theorem, the sequence allows to find the number of distinct real roots of $P(x)$, in a given interval, in terms of the number of changes of signs of the values of the Sturm sequence at the end points of the interval, which can even be taken to be $\pm \infty$.
When applied to the whole real line, it gives the total number of real roots of $P(x)$.
Since $\text{deg}(P^{(i+1)})< \text{deg}(P^{(i)})$ for $0 \leq i < j$, the algorithm terminates and it contains in general $\text{deg}(P)+1$ polynomials. The final polynomial, $P^{(j)}(x)$, is the greatest common divisor of $P(x)$ and its derivative. For instance, suppose that we would like to find the number of distinct real roots of $P(x)$ in the real interval $[x_{i}, x_{f}]$; let $\varsigma(x_i)$ and $\varsigma(x_{f})$ be the number of changes of sign of the Sturm chain at $x_{i}$ and $x_{f}$, respectively; then, Sturm's theorem states that the number of real roots of $P(x)$ in $[x_{i}, x_{f}]$ is simply given by $\varsigma(x_{i})-\varsigma(x_{f})$.

For each value of $\zeta>0$, we have four values of $\lambda$, as $\mathcal{P}(\zeta;\lambda,q,p,r)$ is a $4^{\mathrm{th}}$ degree polynomial in $\lambda$. For all $p$, $r$ real, there always exist four values of $\lambda$ such that $P_{W}(w;\lambda,q,p,r)$ has at least one double $w$-root. The nature of these four points can be classified in terms of the sign of $\Delta_{\lambda}\Delta_{w}P_{W}(w;\lambda,q,p,r)$.
Observe that $\mathcal{P}(0;\lambda,q,p,r)$ coincides with $\Delta_{w}P_{W}(w;\lambda,q,p,r)$, the discriminant of $P_W(w;\lambda,q,p,r)$ with respect to $w$, with reversed sign.
Furthermore, the discriminant of $\mathcal{P}(0;\lambda,q,p,r)$ with respect to $\lambda$, $\Delta_{\lambda}\mathcal{P}(0;\lambda,q,p,r)$, coincides with $\Delta_{\lambda}\Delta_{w}P_{W}(w;\lambda,q,p,r)$. Thus, a first classification of $\mathbf{S}_{x}$ can be achieved by imposing
\[
\Delta_{\lambda}\mathcal{P}(0;\lambda,q,p,r)=-1048576\,(p-r)\,(p+r)\,\left[(p-1)\,(p+8)^{2}-27\,r^{2}\right]^{3}=0\,,
\]
see Proposition \ref{prop:8}.

In order to obtain a complete classification of $\mathbf{S}_{x}$ for all $\zeta$, we observe that, by moving $\zeta$ in the interval $(0,+\infty)$, we move the four $\lambda$-roots of the polynomial $\mathcal{P}(\zeta;\lambda,q,p,r)$ (regarded as a polynomial in $\lambda$).
Two of such roots will collide if
$\Delta_{\lambda}\mathcal{P}(\zeta;\lambda,q,p,r)=\mathcal{Q}(\zeta;q,p,r)\equiv\mathcal{Q}(\zeta)$ vanishes.
Since
\begin{subequations}
\begin{equation}
\label{eq:QQ}
\Delta_{\lambda}\mathcal{P}(\zeta;\lambda,q,p,r)=\mathcal{Q}(\zeta)=65536\,\mathcal{Q}_{1}^{2}(\zeta)\,\mathcal{Q}_{2}(\zeta)\,,
\end{equation}
\begin{align}
\label{eq:q1}
\mathcal{Q}_{1}(\zeta)=&\,p^{3}-27\,r^{2}-3\,p^{2}\,(\zeta-5)-12\,p\,(\zeta-4)+4\,(\zeta-1)\,(\zeta-4)^{2}\,,\\
\nonumber\\
\label{eq:q2}
\mathcal{Q}_{2}(\zeta)=&\,4\,p^{5}\,(\zeta-4)+p^{4}\,(\zeta-4)\,(\zeta+60)+16\,p^{3}\,(3\,\zeta^{2}+r^{2}-48)+\nonumber\\
&+8\,p^{2}\,\left[\zeta\,(\zeta-4)\,(\zeta+36)-32\,(\zeta-4)+r^{2}\,(84-15\,\zeta)\right]+\nonumber\\
&+32\,p\,\left[4\,\zeta\,(\zeta-4)\,(\zeta-2)-3\,r^{2}\,\zeta\,(\zeta-4)+24\,r^{2}\right]+\nonumber\\
&+16\,\left[\zeta^{2}\,(\zeta-4)^{2}-r^{2}\,(\zeta-8)\,(\zeta-2)\,(\zeta+4)-27\,r^{4}\right]\,,
\end{align}
\end{subequations}
and, because $\mathcal{Q}_{1}(\zeta)$ appears squared in the expression of $\mathcal{Q}(\zeta)$, the sign of $\mathcal{Q}(\zeta)$ depends solely on the sign of $\mathcal{Q}_{2}(\zeta)$.
Moreover, we observe that the real, positive roots of $\mathcal{Q}_{1}(\zeta)$ are not associated with any change in the behaviour of $\lambda$, as $\mathcal{Q}(\zeta)$ does not change sign if $\zeta$ passes through one of those roots: indeed, if $\zeta$ passes through one of the roots of $\mathcal{Q}_{1}(\zeta)$, then two $\lambda$-roots will collide, but after the collision they will remain on the branches that they occupied before the collision.

It is easy to see, by isolating the leading term in $\zeta$ of $\mathcal{Q}(\zeta)$, that $\mathcal{Q}(\zeta)>0$ for large $\zeta$. In particular, all the $\lambda$-roots of $\mathcal{P}(\zeta;\lambda,q,p,r)$ are real if $\zeta$ is larger than the maximum of the real, positive roots of $\mathcal{Q}_2(\zeta)$.
After that point there are no complex branches in the $\lambda$-plane, and this provides a simple limit for the largest value of $k_3=\sqrt{\zeta}$ for which the gain function is defined (see (\ref{GAIN})). This also proves that the real axis or part of it is always included in the locus on the $\lambda$-plane.
On the other hand, if $\mathcal{Q}(\zeta)<0$, then two of the $\lambda$ roots are real and two of the $\lambda$-roots form a pair of complex conjugate values.
By changing the value of $\zeta$ and keeping $\mathcal{Q}(\zeta)$ negative, two of the $\lambda$-roots will move on the real axis and the other two will move in the complex plane.
When $\mathcal{Q}(\zeta)=0$, then two of the $\lambda$-roots collide in one point. Finally, if $\mathcal{Q}(\zeta)>0$ for $\zeta>0$ and smaller than the maximum of the real roots of $\mathcal{Q}_2(\zeta)$, we can have either four real roots or two pairs of complex conjugate roots.

Based on what we have seen so far, it is clear that the structure of the algebraic curves in the $\lambda$-plane (\textit{i.e.} the $x$-spectrum $\mathbf{S}_{x}$) is related to the number of changes of sign of $\mathcal{Q}({\zeta})$ for $\zeta\geq0$, that is the number of changes of sign of $\mathcal{Q}_{2}(\zeta)$ for $\zeta\geq0$. In turn, as one varies $p$ and $r$, the number of changes of sign of $\mathcal{Q}_{2}(\zeta)$ for $\zeta\geq0$ can be obtained by counting the number of positive roots of $\mathcal{Q}_{2}(\zeta)$. In particular, as $p$ and $r$ vary, the number of positive roots of $\mathcal{Q}_{2}(\zeta)$ changes in two possible ways: either two roots collide on the real $\zeta$ axis and then move onto the complex plane as a pair of complex conjugate roots (or vice versa), or a positive root passes through the origin and becomes negative (or vice versa). Therefore, in order to provide a complete classification of all the algebraic curves (\textit{i.e.}, the spectra) in the $\lambda$-plane, not only we need to understand the number of changes of sign of $\mathcal{Q}_{2}(\zeta)$ for $\zeta\geq0$ (which is the number of its positive zeros), but also the general structure of the zeros of $\mathcal{Q}_{2}(\zeta)$ (namely, the nature of the its non-positive roots).

To this end, we compute the Sturm chain $\{\mathcal{Q}_2^{(j)}\}_{j=0}^{4}=\{\mathcal{Q}^{(0)}_2,\, \mathcal{Q}^{(1)}_2,\, \mathcal{Q}^{(2)}_2,\,\mathcal{Q}^{(3)}_2,\,\mathcal{Q}^{(4)}_2\}$ of $\mathcal{Q}_{2}(\zeta)$ at $\zeta=0$ and the leading terms of the Sturm chain as $\zeta\to+\infty$, with
\[
\begin{aligned}
& \mathcal{Q}^{(0)}_2(\zeta)=\mathcal{Q}_{2}(\zeta)\,, \\
& \mathcal{Q}^{(1)}_2(\zeta)=\frac{\mathrm{d}}{\mathrm{d}\zeta}\mathcal{Q}^{(0)}_2(\zeta)\,,\\
& \mathcal{Q}^{(2)}_2(\zeta)=-\text{Remainder}\left(\mathcal{Q}^{(0)}_2(\zeta),\mathcal{Q}^{(1)}_2(\zeta)\right)\,,\\
& \mathcal{Q}^{(3)}_2(\zeta)=-\text{Remainder}\left(\mathcal{Q}^{(1)}_2(\zeta),\mathcal{Q}^{(2)}_2(\zeta)\right)\,,\\
& \mathcal{Q}^{(4)}_2(\zeta)=-\text{Remainder}\left(\mathcal{Q}^{(2)}_2(\zeta),\mathcal{Q}^{(3)}_2(\zeta)\right)\,.\\
\end{aligned}
\]
Because of the parametric dependence on $p$ and $r$, the expressions of these quantities are rather large and we prefer not to write the terms of the sequence explicitly here. Just to give the reader an idea, the constant term of the sequence reads
\[
\mathcal{Q}^{(4)}_2(\zeta)=-\frac{\text{Num}(\mathcal{Q}^{(4)}_2)}{\text{Den}(\mathcal{Q}^{(4)}_2)}
\]
with $\text{Num}(\mathcal{Q}^{(4)}_2)$ equal to
\[
r^2 (-p + r) (p + r) (256 + 160 p^2 + p^4 - 12 p^2 r^2 +
      12 r^4)^2 (4096 - 768 p^2 + 48 p^4 - p^6 - 432 p^2 r^2 +
      432 r^4)^3
\]
and $\text{Den}(\mathcal{Q}^{(4)}_2)$ equal to
\[
\begin{aligned}
& 4 (1048576 p^2 - 262144 p^4 + 24576 p^6 -
      1024 p^8 + 16 p^{10} + 4063232 p^2 r^2 + 126976 p^4 r^2+\\
&  -
      8448 p^6 r^2 - 944 p^8 r^2 - p^{10} r^2  4063232 r^4 -
      126976 p^2 r^4 + 77568 p^4 r^4 + 4400 p^6 r^4 + p^8 r^4+ \\
&  -
      138240 p^2 r^6 - 6912 p^4 r^6+ 69120 r^8+ 3456 p^2 r^8)^2\,.
\end{aligned}
\]
If $\varsigma(0)$ is the number of changes of sign for the Sturm chain $\{\mathcal{Q}_2^{(j)}\}_{j=0}^{4}$ at $\zeta=0$ and $\varsigma(\infty)$ the number of changes of sign of the leading terms of the Sturm chain as $\zeta\to+\infty$, then, by Sturm's theorem, the number of positive roots of $\mathcal{Q}_{2}(\zeta)$ is simply given by $\varsigma(0)-\varsigma(\infty)$.
Using this approach, we obtain that, for all $r$, $p$ real, $\mathcal{Q}_{2}(\zeta)$ has always at least one real, positive root. Moreover, in this way we obtain a set of ten algebraic curves in the $(r,p)$-plane, whose intersection defines a set of regions where either $\varsigma(0)$, or $\varsigma(\infty)$, or both change value, namely where the structure of the zeros of $\mathcal{Q}_{2}(\zeta)$ is expected to vary.

Finally, by checking carefully each one of these regions and using some elementary, classical invariant theory \cite{Rees1922},
after a long and tedious analysis, we verify that the following four curves (see (\ref{p_S_implicit}), (\ref{threshpm}), and (\ref{epliC}))
\begin{align*}
& p-r=0\,,\\
& p+r=0\,,\\
& (p-1)\,(p+8)^{2}-27\,r^{2}=0\,,\quad p=p_{S}(r)\,,\\
& (p^{2}-16)^{3}+432\,r^{2}\,\left(p^{2}-r^{2}\right)=0\,,\quad p=p_{C}(r)\,,\quad\mbox{with}\; p<0\,,
\end{align*}
determine the regions where $Q_{2}(\zeta)$ changes the number of positive and negative roots, and hence the structure of its roots. In particular we have the following
 \begin{Proposition}
 \label{prop:11}
 For  $r\geq 0$ and $q=1$, the polynomial $Q_{2}(\zeta;q,r,p)$ has the following structure of zeros.

 If $r < 4$
\begin{center}
\begin{tabular}{ll}
\hline
$- \infty < p < p_C(r)$ & 2 positive, 0 negative, and 2 complex conjugate roots\\
\hline
$ p_C(r) < p < -r$   & 4 positive, 0 negative, and 0 complex conjugate roots\\
\hline
$-r < p < r$  & 1 positive, 1 negative, and 2 complex conjugate roots\\
\hline
$ r < p < p_S(r)$ & 2 positive, 2 negative, and 0 complex conjugate roots\\
\hline
$ p_S(r) < p < -p_C(r)$ & 1 positive, 3 negative, and 0 complex conjugate roots\\
\hline
$ -p_C(r) < p < +\infty$ & 1 positive, 1 negative, and 2 complex conjugate roots\\
\hline
\end{tabular}
\end{center}
 If $r > 4$
\begin{center}
\begin{tabular}{ll}
\hline
$- \infty < p < -r$ & 2 positive, 0 negative, and 2 complex conjugate roots\\
\hline
$ -r < p < p_C(r)$   & 3 positive, 1 negative, and 0 complex conjugate roots\\
\hline
$p_C(r) < p < -p_C(r)$  & 1 positive, 1 negative, and 2 complex conjugate roots\\
\hline
$-p_C(r) < p < p_S(r)$  & 1 positive, 3 negative, and 0 complex conjugate roots\\
\hline
$ p_S(r) < p < r$ & 2 positive, 2 negative, and 0 complex conjugate roots\\
\hline
$ r < p < +\infty$ & 1 positive, 1 negative, and 2 complex conjugate roots\\
\hline
\end{tabular}
\end{center}
At the threshold values, either $Q_{2}(\zeta;q,r,p)$ features a double root, or a positive root passes through the origin.
 \end{Proposition}
It is not difficult to verify that, for $q=1$, when $Q_{2}(\zeta;q,r,p)$ has 2 positive, 0 negative, and 2 complex conjugate roots, then $\mathbf{S}_{x}$ has 0 gaps, 2 branches, and 0 loops (0G 2B 0L); when $Q_{2}(\zeta;q,r,p)$ has 4 positive, 0 negative, and 0 complex conjugate roots, then $\mathbf{S}_{x}$ has 0 gaps, 2 branches, and 1 loop (0G 2B 1L); when $Q_{2}(\zeta;q,r,p)$ has 1 positive, 1 negative, and 2 complex conjugate roots or 1 positive, 3 negative and 0 complex conjugate roots, then $\mathbf{S}_{x}$ has 1 gap, 1 branch, and 0 loops (1G 1B 0L); when $Q_{2}(\zeta;q,r,p)$ has 3 positive, 1 negative, and 0 complex conjugate roots, then $\mathbf{S}_{x}$ has 1 gap, 1 branch, and 1 loop (1G 1B 1L); finally, when $Q_{2}(\zeta;q,r,p)$ has 2 positive, 2 negative, and 0 complex conjugate roots, then $\mathbf{S}_{x}$ has 2 gaps, 0 branches, and 1 loop (2G 0B 1L). Consequently, the four curves (\ref{p_S_implicit}), (\ref{threshpm}), and (\ref{epliC}) determine the regions where the number and the nature of the components of the algebraic curve $\mathbf{S}_{x}$ change. The regions are those described in Proposition \ref{prop:9} and reported in Figure \ref{fig:spectrum} in Section \ref{subsec:branchesandloops}.

\section{Numerical computation of the spectra}
\label{app:C}
In this section, we briefly illustrate how the spectra can be computed numerically.
Let $\hat{\zeta}_j\geq 0$, $j=1,\dots,\hat{n}$, with $1\leq \hat{n}\leq 4$, be the real, positive roots of the polynomial $\mathcal{Q}_2(\zeta)$ (\ref{eq:q2}). We recall that $\mathcal{Q}_2(\zeta)$ has always at least one real, positive root.
Moreover, the non-real part of the spectrum $\mathbf{S}_{x}$ is the locus on the $\lambda$-plane of the $\lambda$-roots of $\mathcal{P}(\zeta;\lambda,q,p,r)$ for $0\leq \zeta \leq \max_{j}\,\hat{\zeta}_j$.
Thus, to numerically compute a spectrum for a given choice of $r$ and $p$, with $q=1$, it suffices to solve $\mathcal{P}(\zeta;\lambda,q,p,r)=0$ for the same choice of the parameters, for a convenient set of values  of $\zeta$ in the interval $[0,\,\max_{j}\,\hat{\zeta}_j]$. These values will be referred to as $\zeta$-nodes. As zeros of $\mathcal{Q}_2(\zeta)$ are zeros of the discriminant of $\mathcal{P}(\zeta;\lambda,q,p,r)$, with respect to $\lambda$, we expect that when $\zeta$ is close to one of the $\hat{\zeta}_j$'s the algebraic curve $\mathbf{S}_{x}$ undergoes rapid changes. This implies that in order to capture all features of the spectrum, and optimize the root-extracting procedure, it is expedient to distribute the $\zeta$-nodes according to a Chebyshev-Gauss-Lobatto distribution \cite{quarteroni2000numerical} between each one of the intervals $[0,\,\hat{\zeta}_1]$, $[\hat{\zeta}_1,\,\hat{\zeta}_2]$,..., $[\hat{\zeta}_{\hat{n}-1},\,\hat{\zeta}_{\hat{n}}]$.

The simultaneous computation of all the roots of all polynomials has been performed via the standard technique of evaluating the eigenvalues of a companion matrix, as per implemented in MATLAB R2017a.
Note also that all spectra have been verified using an objective-function technique over the imaginary part of the differences of the $w_j$'s, computed by solving directly $P_{W}(\zeta;\lambda,q,p,r)=0$.




\end{document}